\documentclass[
  journal=largetwo,
  manuscript=research-article,
  year=2025,
  volume=42,
]{cup-journal}

\usepackage{amsmath}
\usepackage[nopatch]{microtype}
\usepackage{booktabs}
\usepackage{ae,aecompl}

\usepackage{graphicx}	
\usepackage{amsmath}	
\usepackage{amssymb}	

\usepackage{hyperref}
\hypersetup{
	colorlinks=true,
	urlcolor=blue,    
	linkcolor=black,  
	citecolor=blue,   
}

\newcommand\arcsec{\mbox{$^{\prime\prime}$}}%
\newcommand\fdg{\mbox{$.\!\!^\circ$}}%
\def\farcm{%
 \mbox{.\kern -0.7ex\raisebox{.9ex}{\scriptsize$\prime$}}%
}%
\def\farcs{%
 \mbox{%
  \kern  0.13ex.%
  \kern -0.95ex\arcsec%
  \kern -0.1ex%
 }%
}%

\newcommand{\RNum}[1]{\uppercase\expandafter{\romannumeral #1\relax}}

\newcommand\ion[2]{#1$\;${%
\ifx\@currsize\normalsize\small \else
\ifx\@currsize\small\footnotesize \else
\ifx\@currsize\footnotesize\scriptsize \else
\ifx\@currsize\scriptsize\tiny \else
\ifx\@currsize\large\normalsize \else
\ifx\@currsize\Large\large
\fi\fi\fi\fi\fi\fi
\RNum #2}\relax}%

\def\ref@jnl#1{{#1}}%
\newcommand\apjl{{ApJLetters}}
\newcommand\apjs{{ApJS}}
\newcommand\ao{{Appl.~Opt.}}
\newcommand\aap{{A\&A}}

\title{The radio re-brightening of the Type IIb SN\,2001ig}

\author{Roberto Soria}
\affiliation{INAF-Osservatorio Astrofisico di Torino, Strada Osservatorio 20, I-10025 Pino Torinese, Italy}
\alsoaffiliation{Sydney Institute for Astronomy, School of Physics A28, The University of Sydney, Sydney, NSW 2006, Australia}
\email[R. Soria]{roberto.soria@inaf.it}

\author{Thomas D. Russell}
\affiliation{INAF-IASF, via Ugo la Malfa 153, 90146 Palermo, Italy}

\author{Eli Wiston}
\affiliation{Department of Astronomy, University of California, Berkeley, CA 94720-3411, USA}

\author{Siying Cheng}
\affiliation{College of Astronomy and Space Sciences, University of the Chinese Academy of Sciences, Beijing 100049, China}

\author{Raffaella Margutti}
\affiliation{Department of Astronomy, University of California, Berkeley, CA 94720-3411, USA}
\alsoaffiliation{Department of Physics, University of California, 366 Physics North MC 7300, Berkeley, CA 94720, USA}

\author{Kovi Rose}
\affiliation{Sydney Institute for Astronomy, School of Physics A28, The University of Sydney, Sydney, NSW 2006, Australia}
\alsoaffiliation{Australian Telescope National Facility, CSIRO Astronomy and Space Science, PO Box 76, Epping, NSW 1710, Australia}

\author{Stuart Ryder}
\affiliation{School of Mathematical and Physical Sciences, Macquarie University, NSW 2109, Australia}
\alsoaffiliation{Astrophysics and Space Technologies Research Centre, Macquarie University, Sydney, NSW 2109, Australia}

\author{Giacomo Terreran}
\affiliation{Las Cumbres Observatory, 6740 Cortona Drive, Suite 102, Goleta, CA 93117-5575, USA}
\alsoaffiliation{Department of Physics, University of California, Santa Barbara, Broida Hall, Santa Barbara, CA 93106-9530, USA}


\begin{document}

\begin{abstract}
We study the late-time evolution of the compact Type IIb SN\,2001ig in the spiral galaxy NGC\,7424, with new and unpublished archival data from the Australia Telescope Compact Array and the Australian Square Kilometre Array Pathfinder. More than two decades after the SN explosion, its radio luminosity is showing a substantial re-brightening: it is now two orders of magnitude brighter than expected from the standard model of a shock expanding into a uniform circumstellar wind ({\it{i.e.}}, with a density scaling as $R^{-2}$). This suggests that the SN ejecta have reached a denser shell, perhaps compressed by the fast wind of the Wolf-Rayet progenitor or expelled centuries before the final stellar collapse. We model the system parameters (circumstellar density profile, shock velocity, mass loss rate), finding that the denser layer was encountered when the shock reached a distance of $\approx 0.1$ pc; the mass-loss rate of the progenitor immediately before the explosion was $\dot{M}/v_{\rm w} \sim 10^{-7} M_\odot {\mathrm {~yr}}^{-1} {\mathrm {km}}^{-1} {\mathrm {s}}$. 
We compare SN\,2001ig with other SNe that have shown late-time re-brightenings, and highlight the opposite behaviour of some extended Type IIb SNe which show instead a late-time flux cut-off. 
\end{abstract}

\section{Introduction} 
\label{sec:intro}
The radio evolution of core-collapse supernovae (SNe) years to decades after the explosion is a function both of the last stages of evolution of the progenitor star, and of the energy and geometry of the explosion and associated mass ejection. However, the modelling of such processes is hampered by the scarcity of observational constraints; that is, few SNe have been followed in the radio bands both at early and late times. Typical radio light curves are well modelled \citep{chevalier82,weiler86,weiler02,sramek03,chevalier06a,chevalier06b} by an initial flux increase, over the first few weeks or months; this rise is caused by a decrease of the optical depth of absorbing material in the circumstellar medium (CSM), as the synchrotron-emitting blast wave expands. The peak is reached first at shorter wavelengths, then at longer wavelengths. After the source becomes optically thin at a given frequency, its flux density at that frequency declines (as a first approximation) as a power-law. When the emission is optically thin at all frequencies, the spectral index $\alpha$ (defined as $F_\nu \propto \nu^{\alpha}$) has been observed to approach an asymptotically constant, negative value, typically $-1 \lesssim \alpha \lesssim -0.5$ \citep{weiler86,weiler90}.

In recent years, observational studies of a few decades-old SNe have suggested late-time deviations from the power-law decline. In some sources, the flux drops abruptly and exponentially after a few years; the Type IIb SN 1993J \citep{weiler07} in M81 and the Type IIb SN 2001gd \citep{stockdale07} are well-studied examples of such behaviour. A possible interpretation of such an evolution is that the blast wave has overtaken the dense, slow wind from the red-supergiant progenitor and has reached the lower-density interstellar medium. In other (more numerous) sources, instead, a radio re-brightening (sometimes accompanied by X-ray re-brightening) is observed after a few years \citep{wellons12,chevalier17,stroh21,rose24}. Notable examples of radio re-brightenings include the Type II-peculiar SN 1987A \citep{cendes18} in the Large Magellanic Cloud, the Type IIn SN 1996cr \citep{meunier13} in the Circinus galaxy, the Type Ib SN 2014C \citep{margutti17} in NGC\,7331, the Type IIb SN 2003bg in MCG 05-10-15 \citep{rose24} and the Type IIL SN 2018ivc \citep{maeda23} in NGC\,1068.

At least three alternative scenarios have been invoked to explain re-brightenings in decades-old SNe. It could be due to a denser and clumpier circumstellar shell encountered by the expanding shock wave \citep{chevalier98,chevalier17,margutti17}. Alternatively, it could be the signature of an off-axis, relativistic jet that is becoming less collimated over time and is entering our line of sight \citep{paczynski01,granot02}. Thirdly, it could be the emergence of a pulsar wind nebula \citep{slane17} powered by the newborn neutron star.

Monitoring the long-term evolution of the radio emission is particularly important in the case of Type IIb SNe, for which the mass and evolutionary stage of their progenitors is still actively debated \citep{sravan19,sravan20}. 
Type IIb SNe initially show evidence for hydrogen emission that soon subsides, within a few days after explosion. After this epoch the SN develops prominent He features similar to Ib SNe  
\citep{filippenko88,nomoto93,podsiadlowski93,filippenko97,dessart11,gilkis22}. This suggests that Type~IIb progenitor stars have retained only a small layer of hydrogen ($\sim$0.1 M$_\odot$) at the final collapse. The most likely reason for the substantial (but not complete) loss of hydrogen is envelope stripping in a binary system \citep{podsiadlowski93,claeys11,sravan20}. Tentative observational evidence of a surviving binary companion (a B1--B3 supergiant) was indeed proposed for the IIb SN 1993J \citep{fox14}.

The most interesting open question \citep{chevalier10,yoon17,sravan19,sravan20} related to Type IIb SNe is whether their progenitors are extended, such as red or yellow supergiants (easier to detect in pre-explosion optical images: \citealt{kamble16}) or compact, such as stripped stars or Wolf-Rayet stars, similar to Type-Ibc progenitors (not usually detected in pre-explosion images). Extended progenitors have characteristic radii $R_\ast \sim$10$^{13}$ cm and relatively slow wind speeds $v_{\rm w} \sim 10$--100 km s$^{-1}$; compact progenitors have radii $R_\ast \sim$10$^{11}$ cm, with faster wind speeds $v_{\rm w} \sim 500$--1000 km s$^{-1}$. The fastest SN ejecta reach speeds of $\approx$10,000--15,000 km s$^{-1}$ in extended IIb's, but as high as 
$\approx$30,000--50,000 km s$^{-1}$ in compact IIb events, because the ejecta expand into a faster, thinner progenitor wind \citep{chevalier10}. This difference in the ejecta/CSM interaction, in turn, leads to a different evolution of the radio synchrotron emission, especially in the rise-time versus peak-luminosity plane \citep{kamble16,stroh21}. It is also possible that there is a continuum distribution of SN morphologies spanning Ib's, compact IIb's and extended IIb's, rather than separate classes
\citep{chevalier10,horesh13,stroh21}.

Late time radio studies of IIb events can reveal the epoch of hydrogen-envelope removal and the structure of the CSM, and in this way constrain stellar evolution. 
Well-studied representatives of the extended IIb class are SN 1993J and SN 2001gd \citep{chevalier10}, both of which show late-time drops in their radio emission, as mentioned earlier. The Milky Way SN Cassiopeia A was also an extended IIb event, probably with a red supergiant progenitor \citep{krause08}, although a Wolf-Rayet phase just before the explosion has also been suggested \citep{hwang09}. Conversely, SN 2001ig has been interpreted as a compact-progenitor IIb \citep{ryder04,ryder18}. In this paper, we study the late-time radio evolution of SN 2001ig, to test whether and how it differs from that seen in extended IIb SNe.

\section{Early evolution of SN 2001ig}
SN 2001ig was discovered by \cite{evans01}\footnote{The legendary Rev.\ Robert Evans (1937--2022), the most successful visual discoverer of SNe.}
on 2001 December 10.43 UT, in the face-on spiral galaxy NGC\,7424. 
 A likely explosion date is 2001 December 3 (MJD 52246) \citep{ryder04}, based on the early radio light curve evolution. 
 Following \cite{soria24}, we adopt a galaxy distance $d = 10.8$ Mpc (taken as the average between the Hubble Flow distance of 10.1 Mpc and the Tully-Fisher distance of 11.5 Mpc, as reported in the NASA/IPAC Extragalactic Database); this corresponds to a distance modulus of 30.17 mag and a scale of 52 pc arcsec$^{-1}$. The SN position, determined from Australia Telescope Compact Array (ATCA) maps \citep{soria24}, is R.A.(J2000) $= 22^h\,57^m\,30^{s}.74 (\pm 0\farcs04)$, Dec.(J2000) $= -41^{\circ}\,02^{\prime}\,26\farcs35 (\pm 0\farcs05)$.

\begin{figure}[t]
  \begin{center}
    \includegraphics[width=1.0\textwidth]{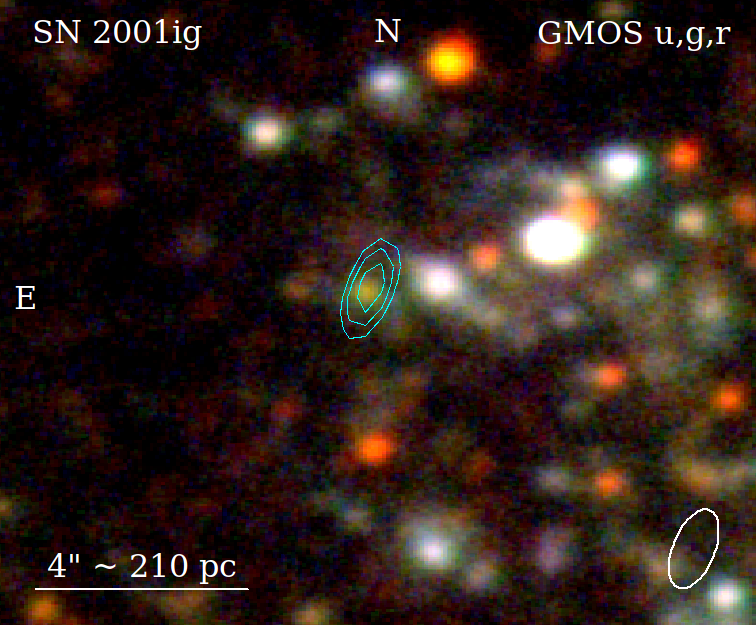}
  \end{center}
  \vspace{-5mm}
  \caption{9 GHz ATCA contours (cyan) of SN\,2001ig observed on 2024 April 24, overlaid on a Gemini/GMOS colour composite image from 2004 September 13. 
  Contours are defined as $2^{n/2}$ times the local rms noise level, which was $\approx$12 $\mu$Jy. The three plotted contours correspond to 68 $\mu$Jy (5.7$\sigma$), 96 $\mu$Jy (8$\sigma$) and 135 $\mu$Jy (11.3$\sigma$). The beam size as shown at lower right was 1\farcs6 $\times$ 0\farcs8 at a position angle (north through east) of 158\fdg0.
  See \cite{ryder06,ryder18} for a discussion on the nature of the optical counterpart.
  }
  \vspace{0.3cm}
\end{figure}

\begin{figure*}[t]
  \begin{center}
\includegraphics[width=0.85\textwidth]{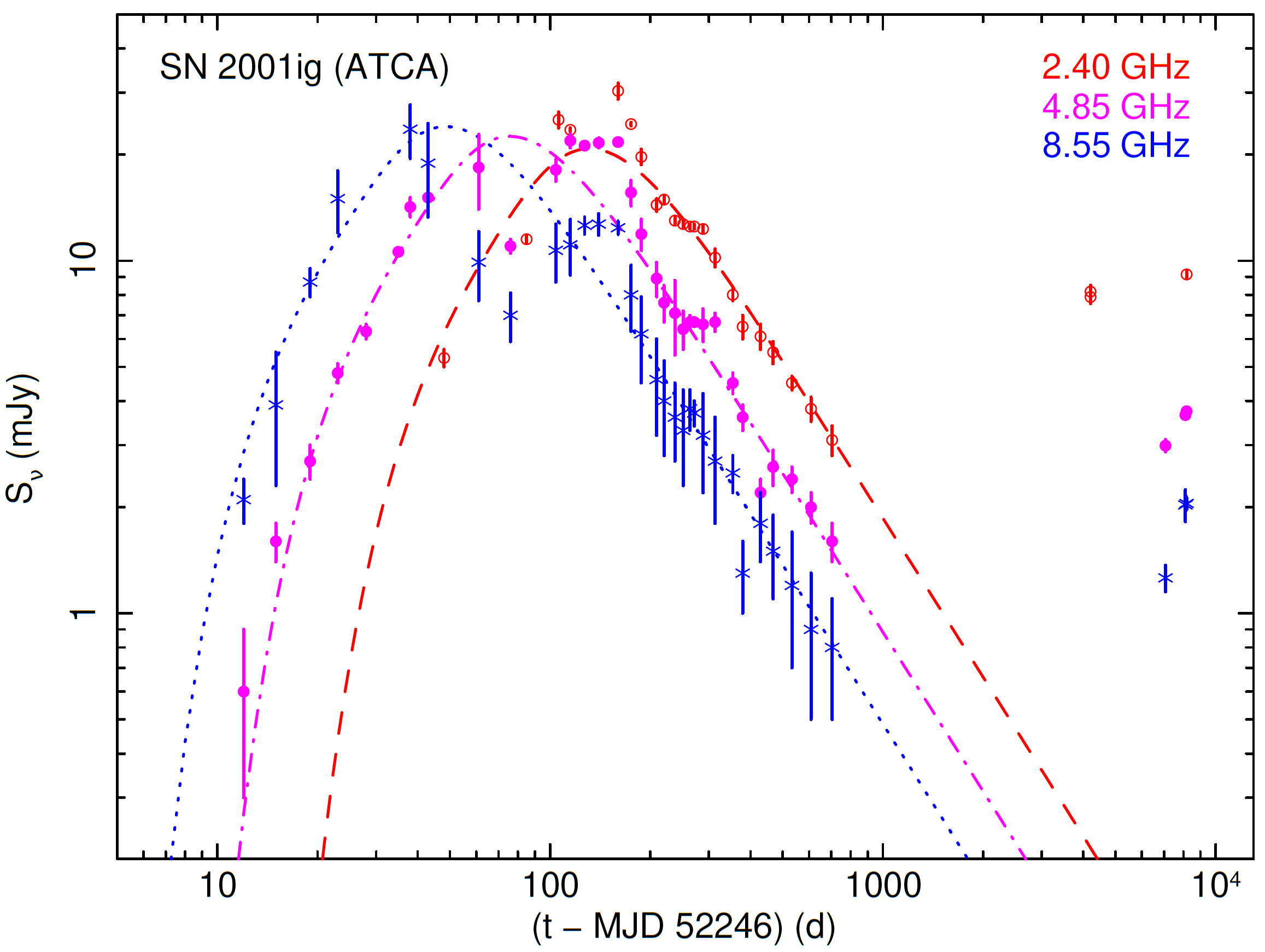}
  \end{center}
  \vspace{-0.2cm}
  \caption{Flux density of SN\,2001ig at 2.4 GHz, 4.85 GHz and 8.55 GHz,
  based on ATCA observations, compared with the canonical evolution model of \cite{weiler02}. The ATCA data from 2001--2003 \citep{ryder04} were taken at central frequencies of 2.4 GHz (red datapoints and dashed line model), 4.85 GHz (magenta datapoints and dash-dotted line model) and 8.55 GHz (blue datapoints and dotted line model), with a bandwidth of 128 MHz. The data from 2013, 2021 and 2024 were taken at central frequencies of 2.1 GHz, 5.5 GHz and 9 GHz (Table 1) and then rescaled to 2.4 GHz, 4.85 GHz and 8.55 GHz with an assumed spectral index $\alpha = -1$, for the purposes of this plot. The bandwidth of the 2013, 2021 and 2024 measurements is 2 GHz.
  }
\end{figure*}

Tentative identification of 2001ig as a Type IIb was already suggested from optical spectra taken from the 6.5-m Baade Magellan Telescope at Las Campanas Observatory on 2001 December 11--13 \citep{phillips01}. The weakening of the H lines and the general similarity with the template Type Ib SN 1993J (but with faster ejecta velocity) was confirmed with spectra taken from the 3.6-m telescope at the European Southern Observatory over the following month \citep{clocchiatti01,clocchiatti02}.
Intensive follow-up optical spectroscopic studies over the first couple of years, with Keck \citep{filippenko02,silverman09}, Gemini-South \citep{ryder06} and the Very Large Telescope  \citep{maund07}, provided comprehensive coverage of the early hydrogen-rich photospheric phase, then of the disappearance of the hydrogen lines, and finally (after $\approx$250 days from the explosion) of the nebular phase, dominated by emission lines such as [\ion{Mg}{1}]\,$\lambda4571$, [\ion{O}{1}]\,$\lambda\lambda6300,6364$, and [\ion{Ca}{2}]\,$\lambda\lambda7291,7324$ (similar to typical Ib SNe). A Gemini/GMOS spectrum at an age of 6 years \citep{ryder18} showed narrow \ion{He}{2}\,$\lambda4686$ emission, similar to that seen to emerge in SN~2014C 1--2 years post-explosion \citep{milisavijevic15}, and interpreted as continuing interaction with a dense CSM. Broad-band optical photometry, based on Gemini/GMOS images from 2004 September 13 (age $\approx$1,000 days), shows a yellow counterpart (Figure 1) consistent with a late-B through late-F supergiant \citep{ryder06}. However, {\it Hubble Space Telescope} Wide Field Camera 3 images in the near-UV (F275W
and F336W filters) from 2016 April 28 (age $\approx$14.4 yrs) showed a fainter, bluer point-like source \citep{ryder18}, consistent with a B2 type ($19,000 {\mathrm K} \lesssim T_{\rm eff} \lesssim 22,000 {\mathrm K}$) main sequence star. The discrepancy is possibly due to residual emission from the shocked gas in the earlier observations, while the later observations are consistent with a $\approx$9 M$_\odot$ companion star. The likely presence of a surviving companion star supports the scenario that hydrogen stripping in the progenitor of SN 2001ig was caused by binary interaction.

%

In the radio bands, high cadence observations were taken over the first $\sim$3 years with the ATCA at 1.4 GHz, 2.4 GHz, 4.8 GHz and 8.6 GHz, with a few, sparser datapoints at 18.8 GHz \citep{ryder04}. In addition, the SN was observed a few times with the Very Large Array (VLA) at 1.4 GHz, 4.9 GHz, 8.5 GHz, 15 GHz and 22.5 GHz \citep{ryder04}. The data showed \citep{ryder04} a general early trend consistent with the model of \citet{weiler02} and \citet{sramek03}, with an initial rise (optically thick phase) followed by an asymptotic power-law decline (optically thin phase), with $F_\nu \propto (t-t_0)^{-1.5}$. In the optically thin phase, the spectral index settles at $\alpha \approx -1$. However, there is an interesting modulation, with a period of $\approx 150$ d,  superposed on the standard flux evolution \citep{ryder04}. This was interpreted as a density modulation of the CSM encountered by the expanding blast wave. One scenario is a series of shell ejections or increased mass loss at regular time intervals over the last few thousand years of the progenitor's life. This may be the mechanism responsible for the dense dust shells seen around the WR-OB binary WR\,112 \citep{lau17} and the recurrent (optical) re-brightenings over the first year of Type IIn supernova iPTF13z  \citep{nyholm17}. Alternatively, \cite{ryder04} suggested that the density enhancements follow a spiral structure (pinwheel or lawn sprinkler model) typical of dust produced by wind-wind collisions near periastron in eccentric massive binary systems. The dusty pinwheel model was theoretically illustrated via hydrodynamical simulations for example by \cite{schwarz96} and \cite{walder03}, and successfully applied to several observed WR-OB systems \citep{tuthill99,monnier99,marchenko02,soulain18}.

\begin{figure}[t]
  \begin{center}
    \includegraphics[width=1.0\textwidth]{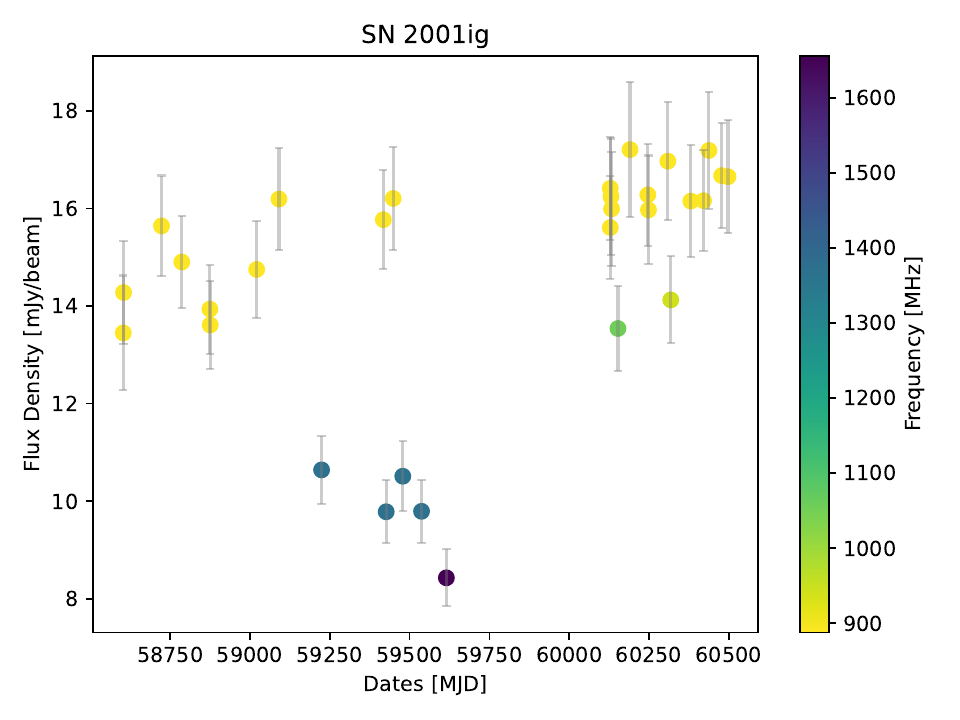}
  \end{center}
  \vspace{-5mm}
  \caption{ASKAP detections of SN 2001ig at late times, starting around $\approx 6,300$\,days. The points are coloured by their central observing frequency, with the majority of the observations coming from the $887.5$\,MHz VAST survey.}
  \vspace{0.3cm}
  \label{fig:askap}
\end{figure}

\begin{table*}[t]
\caption{Summary of the post-2003 ATCA observations of SN 2001ig}
\label{tab:atca}
 \centering
\begin{tabular}{ccccccccc}
\hline\\[-8pt]
 Start Time &  End Time & MJD Range & Project & Array & Frequency & Beam Size & PA & Flux Density\\[2pt]
   &    & & & & (GHz) & & & (mJy) \\[2pt]
 \hline\\[-8pt]

2013-06-08 14:11:50  &  2013-06-09 01:14:00 & 56451.6--56452.1 & C2573 & 6C & 2.1   &  6\farcs0 $\times$ 3\farcs5  & $-$3\fdg3  & 6.90 $\pm$ 0.28\\[2pt]
2013-06-10 16:34:30  &  2013-06-11 01:18:00 & 56453.7--56454.1 & C2573 & 6C & 2.1 & 7\farcs8 $\times$ 3\farcs2  & $+$2\fdg5  & 7.15 $\pm$ 0.29   \\[2pt]      
\hline\\[-8pt]
2021-04-10 17:48:50  & 2021-04-13 05:00:30  & 59314.7--59317.2 & C3421 & 6D & 5.5 & 2\farcs5 $\times$ 1\farcs4 & $+$2\fdg1  & 2.64 $\pm$ 0.11   \\[2pt]
2021-04-10 17:48:50  & 2021-04-13 05:00:30  & 59314.7--59317.2 & C3421 & 6D &   9.0 & 1\farcs5 $\times$ 0\farcs9 & $-$1\fdg6 & 1.2 $\pm$   0.1 \\[2pt]
\hline\\[-8pt]
2024-02-13 21:44:00  &   2024-02-14 09:21:10 & 60353.9--60354.4  & CX550 & 6A &   5.5 &  2\farcs2 $\times$ 1\farcs2 & $+$10\fdg4 &  3.22 $\pm$  0.09\\[2pt]
2024-02-13 21:44:00  &   2024-02-14 09:21:10 & 60353.9--60354.4  & CX550 & 6A &  9.0 &  1\farcs4 $\times$ 0\farcs8 & $+$10\fdg4  & 1.93 $\pm$ 0.20\\[2pt]
\hline\\[-8pt]
2024-04-19 17:27:10  &   2024-04-20 05:01:20 & 60419.7--60420.2 & C3594 & 6A &  2.1 &   4\farcs6 $\times$ 2\farcs7 & $-$0\fdg01 & 8.0 $\pm$   0.2\\[2pt]
2024-04-24 17:14:00  &   2024-04-25 04:56:00 & 60424.7--60425.2 & C3594 & 6A &  5.5 & 2\farcs4 $\times$ 1\farcs1  &  $-$15\fdg9 & 3.30 $\pm$   0.09\\[2pt]
2024-04-24:17:14:00  &   2024-04-25 04:56:00 & 60424.7--60425.2 & C3594 & 6A &   9.0 &   1\farcs6 $\times$ 0\farcs8 &  $-$22\fdg0 &  1.95 $\pm$  0.08 
\\[2pt]
\hline\\
\end{tabular}
\end{table*}

\section{Results of the new radio observations}
To investigate the behaviour of SN 2001ig at ages $>10$ yr, we used several sets of ATCA observations in the 6-km array configuration\footnote{\url{https://www.narrabri.atnf.csiro.au/operations/array\_configurations/configurations.html}.} (Table 1); some of the observations were already in the public archives, while others were specifically obtained for this project. These new data were taken between 2013 June and 2024 April, with 2.1\,GHz data being taken throughout that entire span, and 5.5 and 9\,GHz data being taken from 2021 April until 2024 April, where the 5.5 and 9\,GHz data are recorded simultaneously. All new datasets were taken with a bandwidth of 2 GHz around each central frequency (with the Compact Array Broadband Backend; \citealt{2011MNRAS.416..832W}), compared with the 128 MHz bandwidth of the 2001--2003 data. Details and results from the new ATCA observations are shown in Table~\ref{tab:atca}.

Flux density calibration was done with the primary ATCA calibrator PKS B1934$-$638, and phase calibration with the nearby source PKS B2310$-$417. We processed data following standard procedures\footnote{\url{https://casaguides.nrao.edu/index.php/ATCA\_Tutorials}} within the Common Astronomy Software Application ({\sc casa}, version 5.1.2; \citealt{casa22}). Imaging used the \textsc{casa} task \textit{clean}, using a Briggs robust parameter of 0 \citep{briggs95}, which balances sensitivity and resolution. 
With this choice, we obtained the synthesized beam shapes listed in Table~\ref{tab:atca}.  

To measure the flux density of SN 2001ig, we fit for a point source in the image plane. To do this, we used the {\sc casa} task {\it imfit} to fit a 2-D Gaussian with a full-width-at-half-maximum fixed to the parameters of the synthesized beam. All flux density errors include a systematic uncertainty on the absolute flux density of $2\%$ \citep[e.g.,][]{massardi11,mcconnell12}, which was added in quadrature with the measured noise in a source-free region close to the source position. 

We note that ATCA observations taken on 2024 February 13 suffered from significant phase decorrelation due to poor observing conditions (weather). To address this in the 5.5 GHz data, where there were a number of bright sources in the field, we carried out iterations of phase-only self-calibration, down to a solution interval of 60 seconds (shorter time intervals did not further improve the image). For the 9 GHz data, we were not able to confidently self-calibrate due to the lack of any sufficiently bright sources in the field (all field sources were $<$5 mJy at 9\,GHz). Therefore, to estimate the effects of the phase decorrelation, we re-calibrated the data treating every second scan of the phase calibrator as a 'target' and every other phase calibrator scan as the 'phase calibrator'. Doing so reveals the 'target' phase calibrator scans were fainter by a factor of 0.52, indicating strong phase decorrelation. We then correct for this by applying that correction factor to the initial measurement of SN 2001ig ($1.0 \pm 0.1$ mJy).

We also detected SN 2001ig in a number of Australian SKA Pathfinder \citep[ASKAP;][]{2021PASA...38....9H} observations. These observations were conducted as part of the Rapid ASKAP Continuum Survey \citep[RACS;][]{2021PASA...38...58H} and Variables and Slow Transients \citep[VAST;][]{2021PASA...38...54M} projects, and obtained from the CSIRO ASKAP Science Data Archive\footnote{\href{http://data.csiro.au/}{http://data.csiro.au/}} (CASDA). Both surveys use the \texttt{ASKAPSoft} pipeline \citep{2012SPIE.8500E..0LC} for data reduction and the \texttt{Selavy} algorithm \citep{2012MNRAS.421.3242W} to produce fitted source catalogues.
RACS and VAST observations respectively have $\sim$15\,min and $\sim$12\,min integrations, both with a bandwidth of $288$\,MHz. All epochs of VAST are observed at a central frequency of $887.5$\,MHz. RACS includes epochs observed at central frequencies of $887.5, 943.5,$ and $ 1367.5$\,MHz. The ASKAP detections are shown in Figure \ref{fig:askap}. Flux density errors are calculated as the quadrature sum of the fitted error, RMS, and a $6\%$ flux scale uncertainty.



The main results of the late-time radio observations can be summarized as follows:

i) there is strong re-brightening compared with the extrapolated power-law decline of 2001--2002. Today's radio flux density is two orders of magnitude higher than  we expect assuming a simple power-law model. Low-frequency (2.1 GHz) measurements show (Figure 2) that the re-brightening must have started already before 2013 (age of $\approx$4000 d) and more likely, as early as at an age of $\approx$1000 d.  At the assumed distance of 10.8 Mpc, the luminosity density at 5.5 GHz was $\approx$4.6 $\times 10^{26}$ erg s$^{-1}$ Hz$^{-1}$ in 2004 April.

ii) the re-brightening is still ongoing. There is a slight but significant increase over the past $\approx$2000 d, seen both from ATCA (Figure 2 and Table 1) and from ASKAP (Figure 3). 

iii) the spectral index of the optically thin synchrotron spectrum has remained settled around the canonical value $\alpha \approx -1$, a value it had already attained after $\approx$1 year \citep{ryder04}. Specifically, the ATCA and ASKAP  observations from 2024 April suggest a spectral index $\alpha \approx -0.93 \pm 0.05$ between 0.89 GHz and 9.0 GHz, $\alpha \approx -0.97 \pm 0.05$ between 2.1 GHz and 9.0 GHz, and $\alpha \approx -1.07 \pm 0.05$ between 5.5 GHz and 9.0 GHz. Thus, the re-brightening did not correspond to a change in the spectral slope.

The first hint of an increased CSM interaction leading to a re-brightening was already noted by \cite{ryder18}, from the stack of two Gemini/GMOS spectra taken on 2007 July 18 and November 6 (average age $\approx$2100 d). The combined spectrum showed narrow \ion{He}{2}\,$\lambda4686$ emission, a feature sometimes seen in re-brightening SNe.

\vspace{0.3cm}


\vspace{0.3cm}


\section{Modelling}

We assume that the radio emission is made of an optically thick and an optically thin component \citep{chevalier98}. In the optically thick limit, the flux density, $S_{\nu}$, is proportional to the observing frequency $\nu$ such that $S_{\nu} \propto \nu^{5/2}$, while in the optically thin limit, $S_{\nu} \propto \nu^{-(p-1)/2}$, where $p$, the spectral index of the particle energies, is $> 2$ (typically, $p \sim 3$). The peak flux density at a given time is a function of the shock radius $R_{\rm s}$ and the magnetic field strength $B$ \citep{chevalier98}. From $R_{\rm s}(t)$, we can infer the shock velocity $v_{\rm s}$, the CSM density $\rho_{\rm csm}$ and other quantities of interest for the evolution of the system. 

We model SN 2001ig in two slightly different ways. First, as an initial approximation, we take the observed peak flux density at a given frequency, and assume a value of $p \equiv 3$ and a constant shock velocity to derive $\rho_{\rm csm}$. The main purpose of this model is to estimate $\dot{M}/v_{\rm w}$ of the progenitor star. Then, we use a Markov Chain Monte Carlo (MCMC) model to fit $\rho_{\rm csm}$ at each epoch in order to produce the observed flux density. The main purpose of this model is to estimate the density enhancement encountered by the shock wave at late times.

\subsection{Determining the mass loss rate of the progenitor}

We assume energy equipartition between the magnetic field and the relativistic electrons ($\alpha = 1$ in the formalism of \citealt{chevalier98}), with a 10\% fraction of post-shock energy going into each component: $\epsilon_B = \epsilon_e = 0.1$. We also assume $p \equiv 3$, corresponding to an optically thin synchrotron spectral index $\alpha = -1$, and an emission filling factor (fraction of a spherical volume with outer radius $R_{\rm s}$) $f = 0.5$. Then, the outer shock radius and the magnetic field are obtained from Eqs.(13--14) of \cite{chevalier98}, as a function of the peak flux density $S_{\nu,{\rm p}}$, namely
\begin{eqnarray}
R_{\rm s,p} & = & 8.8 \times 10^{15} \left(\frac{\epsilon_B}{\epsilon_e}\right)^{1/19} 
\left(\frac{f}{0.5}\right)^{-1/19} \left(\frac{S_{\nu,{\rm p}}}{{\mathrm {Jy}}}\right)^{9/19}\nonumber \\
&& \left(\frac{d}{{\mathrm {Mpc}}}\right)^{18/19}  
\left(\frac{\nu}{5{\mathrm {\,GHz}}}\right)^{-1}\ {\mathrm {{cm}}}\\
B_{\rm p} & = & 0.58 \left(\frac{\epsilon_B}{\epsilon_e}\right)^{4/19}\,  
\left(\frac{f}{0.5}\right)^{-4/19}\, \left(\frac{S_{\nu,{\rm p}}}{{\mathrm {Jy}}}\right)^{-2/19}  \nonumber\\ 
&&\left(\frac{d}{{\mathrm {Mpc}}}\right)^{-4/19}\,  
\left(\frac{\nu}{5{\mathrm {\,GHz}}}\right)^{-1}\ {\mathrm G} 
\end{eqnarray}
For $S_{\nu,{\rm p}}$,we take the brightest flux density measurement available at any frequency: $S_{18.8} \approx 43$ mJy beam$^{-1}$, from the ATCA at 18.8 GHz at an age $t-t_0 \approx 28$ d. Thus, the radius of the shocked bubble at that time is $R_{\rm s,p} \approx 5.0 \times 10^{15}$ cm, with a field $B_{\rm p}\approx 1.8$ G. The shock velocity is then simply $v_{\rm s} = R_{\rm s,p}/(t-t_0) \approx 20,000$ km s$^{-1}$.

To derive the mass loss rate from the progenitor, we use the standard assumption (adopted for example in \citealt{rose24}) that the magnetic energy density $B^2/8\pi$ is a fraction $\epsilon_B$ of the post shock energy density: $B^2/8\pi \approx (9/8)\,\epsilon_B \rho_{\rm csm}v_{\rm s}^2$. From Eq.(19) of \cite{chevalier98}, for a constant mass-loss rate in a steady progenitor wind and constant shock velocity, we have:
\begin{eqnarray}
\dot{M} & \approx & 5.2 \times 10^{-6}\, \left( \frac{\epsilon_B}{0.1}\right)^{-1} \left( \frac{B_{\rm p}^2}{1\ {\mathrm G}} \right)^{2} 
\left( \frac{t-t_0}{10\ {\mathrm d}}\right)^{2}\nonumber \\
&&\left( \frac{v_{\rm w}}{1000\ {\mathrm {km~s}}^{-1}}\right)\ \ M_\odot {\mathrm ~yr}^{-1}\nonumber \\
&\approx& 1.3 \times 10^{-4} 
\left( \frac{v_{\rm w}}{1000\ {\mathrm {km~s}}^{-1}}\right)\ \ M_\odot {\mathrm ~yr}^{-1}.
\end{eqnarray} 
At the peak flux density, the CSM density is
\begin{equation}
\rho_{\rm csm} \equiv \frac{\dot{M}}{4\pi R^2_{\rm s,p}v_{\rm w}}  \approx 3 \times 10^{-19}\ \ {\mathrm {g~cm}}^{-3}.
\end{equation} 
Comparing these values with the physical quantities derived in the ASKAP survey of \cite{rose24}, we note that the Type Ic SN 2016coi \citep{grupe16,terreran19} appears the most similar to 2001ig at early times.

\subsection{Determining the late-time density enhancement}

To model the evolution of the synchrotron emission from early to late times, we adopt the thin shell approximation \citep{cox72,castor75,weaver77}, 
and solve, via numerical integration, a simultaneous system of equations for the shock radius and velocity at each epoch, the magnetic field, and the peak flux of the synchrotron radiation that results from shock interactions. Specifically, the thin shell differential equation is 
\begin{equation}
    M_{\rm s} \frac{dv_{\rm s}}{dt} = 4\pi R_{\rm s}^2[\rho_{\textrm{ej}}(v_{\textrm{ej}} - v_{\rm s})^2 - \rho_{\textrm{csm}}v_{\rm s}^2]
\end{equation}
\citep{chevalier17}, where $M_{\rm s}$ is the mass of the thin shell comprising shocked ejecta and shocked CSM at the shock radius $R_{\rm s}$, propagating at the shock velocity $v_{\rm s} = dR_{\rm s}/dt$; $v_{\textrm{ej}}$ is the ejecta velocity at the reverse shock; $\rho_{\textrm{ej}}$ and $\rho_{\textrm{csm}}$ are the density profiles of the shocked ejecta and CSM, respectively. 
$M_{\rm s}$ is computed from the numerical integration of the density profiles. 
For the CSM density, we assume again that the post-shock energy density is proportional to the magnetic energy density 
($\rho_{\rm csm}v_{\rm s}^2 \propto B^2/\epsilon_B$ as in Section 4.2).
For the ejecta density profile we adopt 
\begin{equation}
\rho_{\textrm{ej}} = 
\left\{
    \begin{array}{lr}
        F(t-t_0)^{-3}, &  v_{\textrm{ej}} < v_t\\
        F(t-t_0)^{-3}\left(v_{\textrm{ej}}/v_t\right)^{-n}, & v_{\textrm{ej}} \geq v_t
    \end{array}
\right.
\end{equation}
where 
\begin{equation}
    F = \frac{1}{4\pi n} \frac{[3(n-3)M_{\textrm{ej}}]^{5/2}}{[10(n-2)E]^{3/2}},
\end{equation}
\begin{equation}
    v_t = \biggl[\frac{10(n-5)E}{3(n-3)M_{\textrm{ej}}}\biggl]^{1/2},
\end{equation}
and $M_\textrm{ej}$ and $E$ are the total explosion mass and energy. 
Given the uncertainty on the total ejecta mass, we will solve for the CSM profile for a low end estimate of $M_{ej} = 1 M_\odot$ (approximately the ejecta mass estimated by \cite{silverman09}) and a high end estimate of $M_{ej} = 4 M_\odot$; stripped-envelope SN models by \cite{aguilera23} suggest that the mass range between 1 and 4 $M_\odot$ covers about 80\% of events. 
For the 1$M_\odot$ model, we assume a fixed explosion energy of $E = 10^{51}$ erg, corresponding to an initial ejecta velocity of $v_i \approx 10,000$ km/s.
For the 4$M_\odot$ model, instead, we find that the using the same explosion energy results in an initial ejecta velocity, $v_i \approx 5000$ km/s that is too slow to reproduce the observed early time emission with any density profile. In order to obtain an acceptable flux density fit for this high value of $M_{\textrm{ej}}$, we allow the initial ejecta velocity to be a free parameter, $v_{\textrm{i}}$. For this higher mass scenario we find that an initial ejecta velocity of $v_i = 13,888$ km/s is needed to replicate the early time emission.  
Finally, we adopt an ejecta density index $n = 10$, suitable for massive stars with radiative envelopes and convective cores \citep{chevalier82a,matzner99,chevalier17}, including Wolf-Rayet stars \citep{kippenhahn90}.


In addition to Eq.(5) for the shock dynamics, and the equations for ejecta and CSM density, we simultaneously solve Eqs.(11--12) from \cite{chevalier98} to infer the peak flux density 
$S_{\nu,{\rm p}}$ and peak frequency $\nu_{\rm p}$ from the calculated radius and magnetic field strength at each epoch. As we did in Section 4.1, we assume that the fractions of post-shock energy transferred to the magnetic field and the relativistic electrons are $\epsilon_B = 0.1$  and $\epsilon_e = 0.1$, respectively, and that these fractions do not evolve with time; we assume again a standard filling factor, $f = 0.5$. We then model the radio spectrum as a broken power law, with an optically thick slope of 5/2 and an optically thin slope of $(1-p)/2$:
\begin{equation}
    S_\nu = 2 S_{\nu,{\rm p}} \left[ \left(\frac{\nu}{\nu_{\textrm{p}}}\right)^{-(1-p)/2} + \left(\frac{\nu}{\nu_{\textrm{p}}}\right)^{-5/2}\right]^{-1}.
\end{equation}
Unlike what we did in Section 4.1, here we do not assume a spectral index $p=3$ for the electron energies; instead, we fit our broken power-law spectrum across all epochs with an MCMC fit (with the python package {\sc {emcee}}\footnote{\url{https://emcee.readthedocs.io/en/stable}.}),
and find that the best fitting value is constrained to be $p = 2.83 \pm 0.02$.



We assume that at early times, $(t-t_0) < 700$\,d, the SN shock is interacting with a CSM density profile that scales as $\rho_{\textrm{csm}}\propto R^{-\alpha}$. Using an MCMC to model the shock dynamics and resulting radiation for just the early time data, we infer $\alpha = 1.96\pm 0.01$. We thus assume a standard wind density profile ($\rho_{\textrm{csm}} \propto R^{-2}$) at smaller radii. Our main objective here is to explain the observed radio re-brightening at late times (Figure 2). We want to test whether a deviation from the CSM wind-density profile at large radii can explain the re-brightening, and if so, what amount of density enhancement may be needed. To this aim, we introduce a break radius $R_{\rm{brk}}$, where the shock encounters an over-density compared to the extrapolation of the wind-density profile, and starts to interact with a CSM characterized by shallower density index, $\alpha_{\rm{over}}$. For a detailed description of the model, in the more general case of a three-zone CSM, see \cite{idik24}. 

\begin{figure}[t]
  \begin{center}
    \includegraphics[width=1.0\textwidth]{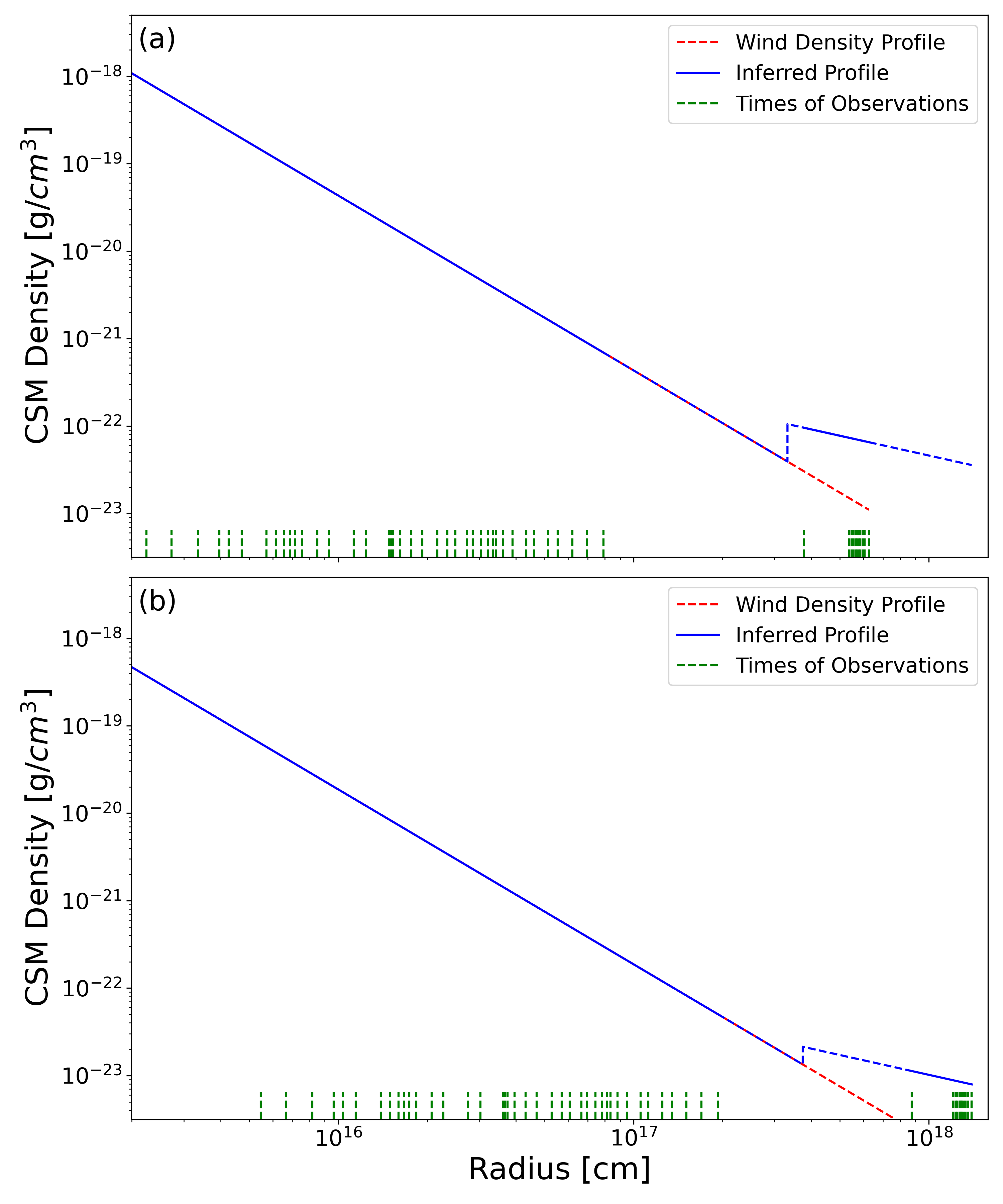}
  \end{center}
  \vspace{-7mm}
  \caption{Inferred CSM density profiles as a function of shock radius, for (a) $M_{\rm{ej}} = 1 M_\odot$, and (b) $M_{\rm{ej}} = 4 M_\odot$. In each panel, the blue line is our model density profile that provides the best fit to the observed radio datapoints at early and late times; the dashed red line is the density profile assuming a uniform stellar wind ($\rho_{\textrm{csm}} \propto R^{-2}$). The green markers on the X axis represent the best-fitting shock radius at each of the epochs with radio observations. The gap in the radio coverage between ages of $\approx$2--12 years creates an uncertainty about where the density enhancement occurs ({\it{i.e.}}, the parameter $R_{\textrm{brk}}$ in our model, Section 4.2). The portions of the CSM density profiles well constrained by the radio observations are plotted as solid blue segments, while the intervals without a strong constraint are dashed.} 
  \vspace{0.3cm}
  \label{fig:density_profile}
\end{figure}

\begin{figure}[t]
  \begin{center}
    \includegraphics[width=1.0\textwidth]{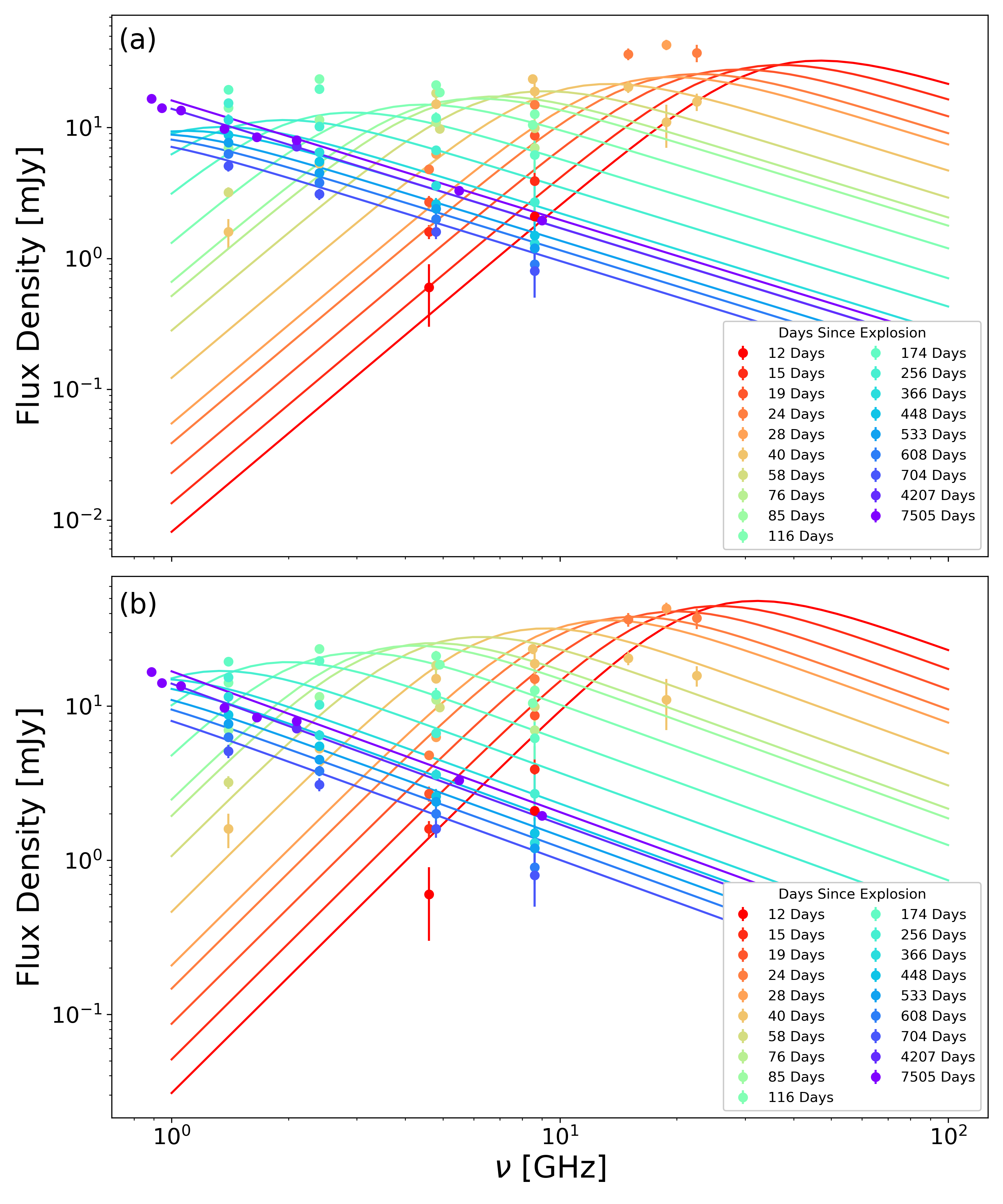}  \\  
  \end{center}
  \vspace{-7mm}
  \caption{Spectral energy distributions of the best-fitting models for (a) $M_{\rm{ej}} = 1 M_\odot$, and (b)  $M_{\rm{ej}} = 4 M_\odot$, at each time step. Datapoints (solid circles) are the observed radio flux densities, colour-coded by SN age, from red (earliest) to violet (latest). Consecutive observations with $\Delta t/t < 0.1$ are combined, and their plotted colour corresponds to the average age. Solid lines are model radio spectra (also colour-coded by age) corresponding to our best-fitting CSM density profile and shock velocity. At earlier times, flux density and peak frequency decrease over time, as expected from a slowly decelerating shock propagating through a uniformly expanding wind. At later times, the flux increases again, which we interpret as a flattening of the CSM density profile.
  }
  \vspace{0.3cm}
  \label{fig:fit-SEDs}
\end{figure}

\begin{figure}[t]
  \begin{center}
    \includegraphics[width=1.0\textwidth]{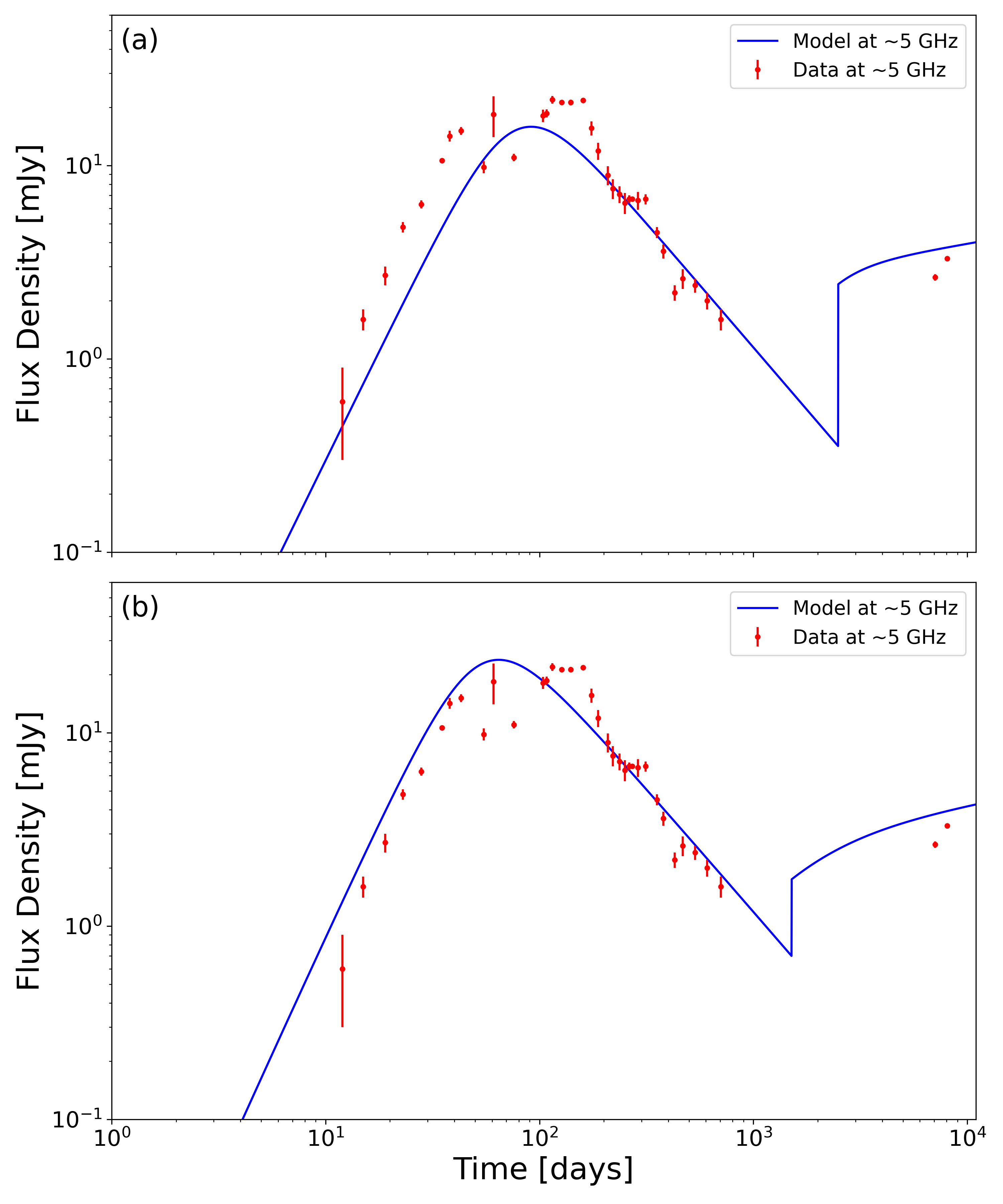}  \\  
  \end{center}
  \vspace{-7mm}
  \caption{Flux density evolution of the best-fitting spectral model for (a) $M_{\rm{ej}} = 1 M_\odot$, and (b)  $M_{\rm{ej}} = 4 M_\odot$, for the representative 5-GHz case, compared with the observed datapoints (red circles). }
  
  \vspace{0.3cm}
  \label{fig:lightcurve}
\end{figure}

We use another MCMC fit to solve for the CSM profile that would produce the observed emission. The fit parameters are: $R_{\textrm{brk}}$, $\rho_{\textrm{0,wind}}$, $\rho_{\textrm{0,over}}$, and $\alpha_{\textrm{over}}$. $R_{\textrm{brk}}$ represents the radius at which the shock encounters an overdensity; $\rho_{\textrm{0,wind}}$ and $\rho_{\textrm{0,over}}$ are the CSM densities immediately before and after $R_{\textrm{brk}}$. We assume that the CSM follows a wind density profile ($\rho_{\textrm{csm}} \propto R^{-2}$) at $R < R_{\textrm{brk}}$ and a shallower profile $\rho_{\textrm{csm}} \propto R^{-\alpha_{\textrm{over}}}$ at $R > R_{\textrm{brk}}$:
\begin{equation}
\rho_{\textrm{CSM}} = 
\left\{
    \begin{array}{lr}
        \rho_{\rm 0,wind} \left(R/R_{\rm brk}\right)^{-2}, &  R < R_{\rm brk}\\
        \rho_{\rm 0,over} \left(R/R_{\rm brk}\right)^{-\alpha_{\rm over
        }}, &  R > R_{\rm brk}
    \end{array}
\right.
\end{equation}


For this treatment of the shock dynamics, we expect the shock velocity to decrease steadily over time, as the shock propagates through power law CSM profiles. At the density discontinuity $R_{\rm brk}$, the rate of deceleration will change as the shock encounters an overdensity, with a potentially different power law index, $\alpha_{\rm over}$. With this evolution, we expect the overall flux density to steadily decline as the shock velocity decreases along the wind density section of the CSM profile. When encountering the overdensity at $R_{\rm brk}$, the flux density will increase as the shock accelerates a larger population of electrons to produce synchrotron emission. At very late times ($(t-t_0) \gtrsim 2 \times 10^{4}$\,d and $(t-t_0) \gtrsim 5 \times 10^{4}$\,d for the low and high $M_{\textrm{ej}}$ scenarios, respectively), the flux density will once again begin to decline, as the shock velocity keeps decreasing.

The best-fitting CSM density profiles for the low and high $M_{\textrm{ej}}$ scenarios are shown in Figure \ref{fig:density_profile}, and the corresponding time evolutions of the model flux densities are plotted in Figure \ref{fig:fit-SEDs} and (for the representative 5-GHz case) Figure \ref{fig:lightcurve}. Details of the fit results for each parameter are in the Appendix, Figures \ref{fig:corner_plot1}--\ref{fig:corner_plot2}. Although the density jump at the break radius is somewhat small in both cases, this increase, in combination with the shallower power-law index $\alpha_{\rm{over}}$, creates an overdensity of about an order of magnitude at the most recent epochs of observation (year 2024) compared to what it would have been in case of a standard wind profile. This, in turn, leads to radio flux densities about two orders of magnitude higher than expected from the extrapolation of the early-time decline. The best fit value for $\alpha_{\rm{over}}$ remains constant across both fits, at $\alpha_{\rm{over}} = 0.75$. This shallow index, corresponding to a slow evolution of CSM density, is needed to reproduce the slow evolution of synchrotron emission after the re-brightening. 

As a caveat, we stress that the lack of data between $\approx$3--10 years post-explosion creates uncertainty as to where exactly the overdensity begins. This can clearly be seen from the large gap between the fitted values of the shock radii at each epoch of observations, marked as green segments on the X axis of Figure \ref{fig:density_profile}; the overdensity begins somewhere within that gap. As a result, although our best-fitting parameters represent one solution in good agreement with the observations, there are still strong degeneracies between $R_{\textrm{brk}}$ and the normalizations of $\rho_{\textrm{0,wind}}$ and $\rho_{\textrm{0,over}}$. A different combination of break radius and overdensity normalization could produce the same densities at the points where we actually have measurements.

The best-fitting density profiles in Figure \ref{fig:density_profile} reach fairly low densities (number densities $n_e \sim 10$--100 cm$^{-3}$), especially in the case of high $M_{\textrm{ej}}$. However, this is a direct result of our choice of the microphysical parameter $\epsilon_B = 0.1$. Some models of non-relativistic shock acceleration predict values of $\epsilon_B$ an order of magnitude lower \citep{Gupta24}. 
If we were to assume a lower fraction of the total energy transferred to the magnetic field strength, we would in turn infer higher densities, to produce the same observed radio emission. Therefore, our density models can best be interpreted as lower limits on the CSM density profile. 
While the density normalization remains uncertain, the inferred trend of a flattening density index at late times is a solid result regardless of the choices of $M_{\textrm{ej}}$ and $\epsilon_B$.

\section{Discussion and conclusion}

With new ATCA and ASKAP observations two decades after the event, we report a radio re-brightening of SN 2001ig. The remnant is now two orders of magnitude brighter than expected from the extrapolation of its initial power-law decline, which was well monitored during the first 2 years. The increase in radio brightness is still ongoing; the most recent measurement of the luminosity density at 5.5 GHz was $\approx$4.6 $\times 10^{26}$ erg s$^{-1}$ Hz$^{-1}$ in 2004 April. Despite the radio flux changes, the spectral index has remained constant at $\alpha \approx -1.0$, typical of the optically thin emission expected in radio SNe. This finding suggests that the re-brightening is caused by the shock wave passing through a denser CSM region; in fact, the ambient density encountered by the shock may still be declining with radius, but slower than the canonical $R^{-2}$ wind profile. Other scenarios for the re-brightening are more contrived or ruled out. For example, in the young pulsar wind nebula scenario, a flattening of the radio spectral index would have been expected. The off-axis jet scenario is also disfavoured by our data owing to the shape of the radio light-curve, with a peak at early times, a decline, and a re-brightening two decades later (see the discussion of this scenario and its shortcomings in \citealt{stroh21}). At the current age, the CSM over-density required to explain the flux increase is about an order of magnitude above the extrapolated wind density profile. From our modelling of the radio flux density evolution at early and late times, we estimate that the denser layer was encountered when the shock reached a distance of $\approx$3 $\times 10^{17}$ cm $\approx 0.1$ pc. 
The mass-loss rate of the progenitor immediately before the explosion is estimated at $\dot{M}/v_{\rm w} \approx 1.3 \times 10^{-7} M_\odot {\mathrm {~yr}}^{-1} {\mathrm {km}}^{-1} {\mathrm {s}}$.
We stress that this solution is based on a simple two-zone structure of the CSM, with a density jump between them, justified by the limited number of late-time datapoints available so far. Alternative solutions based on more complex density structures, for example with the presence of a finite-width, high-density wall between the ejecta and the inner part of the CSM \citep{harris20}, are left to follow-up work.

The re-brightening of Type IIb SN 2001ig differs from the behaviour shown by the well-studied Type IIb SN 1993J, whose late-time radio flux dropped sharply below the extrapolated decline model. Such behaviour is consistent with the interpretation of Type IIb SNe as a mixed class, some with extended red/yellow supergiant progenitors (slower winds and late-time radio downturn) and others with compact Wolf-Rayet or stripped progenitors (late-time re-brightening). SN 2001ig joins the small but growing sample of re-brightening SNe, previously identified from the VLA Sky Survey \citep{stroh21} and from the ASKAP Variables and Slow Transients survey \citep{rose24}.  Its proximity to us makes it one of the best such sources for follow-up studies.

Type IIb SN 2003bg ($d \approx 19.5$ Mpc) is one of the best analogues of 2001ig \citep{soderberg06,rose24}. Both have shown recurrent flux modulations at early times, and re-brightenings at late times. Both  are consistent with a Wolf-Rayet progenitor with mass loss rates $\dot{M}\sim 10^{-4} M_\odot {\mathrm ~yr}^{-1}$. However, the shock velocity in 2003bg is estimated to be $\approx$58,000 km s$^{-1}$ \citep{rose24}, significantly higher than the plausible range of shock velocities we inferred for 2001ig ($v_{\mathrm{s}} \sim 10,000$--20,000 km s$^{-1}$ depending on the model assumptions). Moreover, the peak luminosity of 2003bg is an order of magnitude higher, and the CSM density at peak luminosity is an order of magnitude lower \citep{rose24}. Type Ic SN 2016coi is a close analog of 2001ig in terms of $v_{\mathrm{sh}}$, early-time $\rho_{\mathrm{csm}}$, and $\dot{M}/v_{\mathrm{w}}$ \citep{rose24}. However, 2016coi is thought to come from a single stellar progenitor rather than a binary \citep{terreran19} and does not show significant late-time radio re-brightening \citep{rose24}.

The simplest explanation for late-time re-brightenings is that SN 2001ig exploded in a low-density bubble surrounded by a denser shell with an inner wall at $R \sim v_{\rm s}\, \delta t \approx 0.06\ (v_{\rm s}/2 \times 10^4\ {\rm km~s}^{-1})\,(\delta t/1000\ {\rm d})$ pc, where $\delta t$ is the age at which the shock reached the denser shell. Type Ib/IIn SN 2014C arguably provides the best observational evidence (from radio to hard X-rays) of an interaction of the fast SN shock with the slower, denser shell of $\approx$1 $M_\odot$ of hydrogen-rich material at $\approx$6 $\times 10^{16}$ cm $\approx$0.02 pc \citep{margutti17,brethauer20}. 

The physical origin of the denser shell is still unclear, and may differ from system to system. A simple explanation is that a shell can be interstellar gas and/or slower wind from an earlier red supergiant phase of the progenitor or of the binary companion, swept-up by the faster Wolf-Rayet wind in the centuries before the explosion. An alternative possibility is that the shell was part of the envelope of the progenitor star, ejected via some kind of internal instability. In the case of SN 2001ig, the progenitor star is not massive enough to be a Luminous Blue Variable or $\eta$-Carinae-like star, in which shell ejections are well known; however, other forms of instability may exist. For example, \cite{quataert12} proposed that convection inside C-fusing stars can drive internal gravity waves, which are then converted to acoustic waves. Such waves carry a super-Eddington amount of power; their dissipation in the stellar envelope can unbind up to several $M_\odot$ of H-rich material and trigger massive outflows. Another intriguing scenario \citep{margutti17} is that the presence of a denser shell, and therefore of a late-time re-brightening, is a signature of a binary progenitor system. Specifically, binary interactions may cause runaway mass transfer, followed by a common envelope, then by the ejection of the common envelope at low speeds. The leftover stripped star is the one that explodes as a SN. Finally, the fast ejecta interact with the slow expanding common envelope.

We plan to continue regular radio monitoring of SN 2001ig over the coming years, aiming to determine the thickness and total mass of the interacting CSM shell, and hence constrain its formation scenarios. In addition, if the thick shell was part of the progenitor envelope or a common envelope, we expect a re-appearance or re-brightening of Balmer optical emission lines. This is indeed the case for example in the Type Ib SN 2004dk, whose radio re-brightening (starting at an age of $\approx$1700 d) has been well monitored over several years \citep{wellons12,balasubramian21,rose24}. SN 2004dk originally showed a He-rich. H-poor spectrum; however, broad H$\alpha$ emission, due to the interaction between SN shock and denser CSM, was visible from an age of $\approx$3500 d \citep{vinko17,pooley19}.

\begin{acknowledgement}

We thank Laura Driessen, Ashna Gulati, Andr\'es Gurpide, Matt Middleton, Lida Oskinova, Manfred Pakull for discussions and suggestions. RS acknowledges support and hospitality from the School of Physics, University of Sydney, and from the National Astronomical Observatories of China (Beijing) during part of this work. 
TDR is supported by a IAF-INAF Research Fellowship.
KR thanks the LSST-DA Data Science Fellowship Program, which is funded by LSST-DA, the Brinson Foundation, and the Moore Foundation; their participation in the program has benefited this work.
The Australia Telescope Compact Array is part of the Australia Telescope National Facility ({\url {https://ror.org/05qajvd42}}) which is funded by the Australian Government for operation as a National Facility managed by CSIRO. We acknowledge the Gomeroi people as the Traditional Owners of the ATCA observatory site.
This scientific work uses data obtained from Inyarrimanha Ilgari Bundara / the Murchison Radio-astronomy Observatory. We acknowledge the Wajarri Yamaji People as the traditional owners of the Observatory site. CSIRO’s ASKAP radio telescope is part of the \href{https://ror.org/05qajvd42}{Australia Telescope National Facility}. Operation of ASKAP is funded by the Australian Government with support from the National Collaborative Research Infrastructure Strategy. ASKAP uses the resources of the Pawsey Supercomputing Research Centre. Establishment of ASKAP, Inyarrimanha Ilgari Bundara, the CSIRO Murchison Radio-astronomy Observatory and the Pawsey Supercomputing Research Centre are initiatives of the Australian Government, with support from the Government of Western Australia and the Science and Industry Endowment Fund.
\end{acknowledgement}

\paragraph{Funding Statement}

This work was performed in part at the Aspen Center for Physics, which is supported by National Science Foundation grant PHY-2210452.
RS acknowledges the INAF grant number 1.05.23.04.04.

\paragraph{Competing Interests}

None.

\paragraph{Data Availability Statement}

This paper includes archived data obtained through the CSIRO ASKAP Science Data Archive, CASDA\footnote{\href{http://data.csiro.au/}{http://data.csiro.au/}}.

\printendnotes

\printbibliography

@ARTICLE{aguilera23,
       author = {{Aguilera-Dena}, David R. and {M{\"u}ller}, Bernhard and {Antoniadis}, John and {Langer}, Norbert and {Dessart}, Luc and {Vigna-G{\'o}mez}, Alejandro and {Yoon}, Sung-Chul},
        title = "{Stripped-envelope stars in different metallicity environments. II. Type I supernovae and compact remnants}",
      journal = {\aap},
         year = 2023,
        month = mar,
       volume = {671},
          eid = {A134},
        pages = {A134},
          doi = {10.1051/0004-6361/202243519},
archivePrefix = {arXiv},
       eprint = {2204.00025},
 primaryClass = {astro-ph.SR},
       adsurl = {https://ui.adsabs.harvard.edu/abs/2023A&A...671A.134A}
}

@ARTICLE{2011MNRAS.416..832W,
       author = {{Wilson}, Warwick E. and {Ferris}, R.~H. and {Axtens}, P. and {Brown}, A. and {Davis}, E. and {Hampson}, G. and {Leach}, M. and {Roberts}, P. and {Saunders}, S. and {Koribalski}, B.~S. and {Caswell}, J.~L. and {Lenc}, E. and {Stevens}, J. and {Voronkov}, M.~A. and {Wieringa}, M.~H. and {Brooks}, K. and {Edwards}, P.~G. and {Ekers}, R.~D. and {Emonts}, B. and {Hindson}, L. and {Johnston}, S. and {Maddison}, S.~T. and {Mahony}, E.~K. and {Malu}, S.~S. and {Massardi}, M. and {Mao}, M.~Y. and {McConnell}, D. and {Norris}, R.~P. and {Schnitzeler}, D. and {Subrahmanyan}, R. and {Urquhart}, J.~S. and {Thompson}, M.~A. and {Wark}, R.~M.},
        title = "{The Australia Telescope Compact Array Broad-band Backend: description and first results}",
      journal = {\mnras},
         year = 2011,
        month = sep,
       volume = {416},
       number = {2},
        pages = {832-856},
          doi = {10.1111/j.1365-2966.2011.19054.x},
archivePrefix = {arXiv},
       eprint = {1105.3532},
 primaryClass = {astro-ph.IM},
       adsurl = {https://ui.adsabs.harvard.edu/abs/2011MNRAS.416..832W}
}

@ARTICLE{2021PASA...38....9H,
       author = {{Hotan}, A.~W. and {Bunton}, J.~D. and {Chippendale}, A.~P. and {Whiting}, M. and {Tuthill}, J. and {Moss}, V.~A. and {McConnell}, D. and {Amy}, S.~W. and {Huynh}, M.~T. and {Allison}, J.~R. and {Anderson}, C.~S. and {Bannister}, K.~W. and {Bastholm}, E. and {Beresford}, R. and {Bock}, D.~C. -J. and {Bolton}, R. and {Chapman}, J.~M. and {Chow}, K. and {Collier}, J.~D. and {Cooray}, F.~R. and {Cornwell}, T.~J. and {Diamond}, P.~J. and {Edwards}, P.~G. and {Feain}, I.~J. and {Franzen}, T.~M.~O. and {George}, D. and {Gupta}, N. and {Hampson}, G.~A. and {Harvey-Smith}, L. and {Hayman}, D.~B. and {Heywood}, I. and {Jacka}, C. and {Jackson}, C.~A. and {Jackson}, S. and {Jeganathan}, K. and {Johnston}, S. and {Kesteven}, M. and {Kleiner}, D. and {Koribalski}, B.~S. and {Lee-Waddell}, K. and {Lenc}, E. and {Lensson}, E.~S. and {Mackay}, S. and {Mahony}, E.~K. and {McClure-Griffiths}, N.~M. and {McConigley}, R. and {Mirtschin}, P. and {Ng}, A.~K. and {Norris}, R.~P. and {Pearce}, S.~E. and {Phillips}, C. and {Pilawa}, M.~A. and {Raja}, W. and {Reynolds}, J.~E. and {Roberts}, P. and {Roxby}, D.~N. and {Sadler}, E.~M. and {Shields}, M. and {Schinckel}, A.~E.~T. and {Serra}, P. and {Shaw}, R.~D. and {Sweetnam}, T. and {Troup}, E.~R. and {Tzioumis}, A. and {Voronkov}, M.~A. and {Westmeier}, T.},
        title = "{Australian square kilometre array pathfinder: I. system description}",
      journal = {\pasa},
         year = 2021,
        month = mar,
       volume = {38},
          eid = {e009},
        pages = {e009},
          doi = {10.1017/pasa.2021.1},
archivePrefix = {arXiv},
       eprint = {2102.01870},
 primaryClass = {astro-ph.IM},
       adsurl = {https://ui.adsabs.harvard.edu/abs/2021PASA...38....9H}
}

@ARTICLE{2012MNRAS.421.3242W,
       author = {{Whiting}, Matthew T.},
        title = "{DUCHAMP: a 3D source finder for spectral-line data}",
      journal = {\mnras},
         year = 2012,
        month = apr,
       volume = {421},
       number = {4},
        pages = {3242-3256},
          doi = {10.1111/j.1365-2966.2012.20548.x},
archivePrefix = {arXiv},
       eprint = {1201.2710},
 primaryClass = {astro-ph.IM},
       adsurl = {https://ui.adsabs.harvard.edu/abs/2012MNRAS.421.3242W}
}

@INPROCEEDINGS{2012SPIE.8500E..0LC,
       author = {{Cornwell}, T.~J. and {Voronkov}, M.~A. and {Humphreys}, B.},
        title = "{Wide field imaging for the square kilometre array}",
    booktitle = {Image Reconstruction from Incomplete Data VII},
         year = 2012,
       editor = {{Bones}, Philip J. and {Fiddy}, Michael A. and {Millane}, Rick P.},
       series = {Society of Photo-Optical Instrumentation Engineers (SPIE) Conference Series},
       volume = {8500},
        month = oct,
          eid = {85000L},
        pages = {85000L},
          doi = {10.1117/12.929336},
archivePrefix = {arXiv},
       eprint = {1207.5861},
 primaryClass = {astro-ph.IM},
       adsurl = {https://ui.adsabs.harvard.edu/abs/2012SPIE.8500E..0LC}
}

@ARTICLE{2021PASA...38...58H,
       author = {{Hale}, Catherine L. and {McConnell}, D. and {Thomson}, A.~J.~M. and {Lenc}, E. and {Heald}, G.~H. and {Hotan}, A.~W. and {Leung}, J.~K. and {Moss}, V.~A. and {Murphy}, T. and {Pritchard}, J. and {Sadler}, E.~M. and {Stewart}, A.~J. and {Whiting}, M.~T.},
        title = "{The Rapid ASKAP Continuum Survey Paper II: First Stokes I Source Catalogue Data Release}",
      journal = {\pasa},
         year = 2021,
        month = nov,
       volume = {38},
          eid = {e058},
        pages = {e058},
          doi = {10.1017/pasa.2021.47},
archivePrefix = {arXiv},
       eprint = {2109.00956},
 primaryClass = {astro-ph.GA},
       adsurl = {https://ui.adsabs.harvard.edu/abs/2021PASA...38...58H}
}

@ARTICLE{2021PASA...38...54M,
       author = {{Murphy}, Tara and {Kaplan}, David L. and {Stewart}, Adam J. and {O'Brien}, Andrew and {Lenc}, Emil and {Pintaldi}, Sergio and {Pritchard}, Joshua and {Dobie}, Dougal and {Fox}, Archibald and {Leung}, James K. and {An}, Tao and {Bell}, Martin E. and {Broderick}, Jess W. and {Chatterjee}, Shami and {Dai}, Shi and {d'Antonio}, Daniele and {Doyle}, Gerry and {Gaensler}, B.~M. and {Heald}, George and {Horesh}, Assaf and {Jones}, Megan L. and {McConnell}, David and {Moss}, Vanessa A. and {Raja}, Wasim and {Ramsay}, Gavin and {Ryder}, Stuart and {Sadler}, Elaine M. and {Sivakoff}, Gregory R. and {Wang}, Yuanming and {Wang}, Ziteng and {Wheatland}, Michael S. and {Whiting}, Matthew and {Allison}, James R. and {Anderson}, C.~S. and {Ball}, Lewis and {Bannister}, K. and {Bock}, D.~C. -J. and {Bolton}, R. and {Bunton}, J.~D. and {Chekkala}, R. and {Chippendale}, A.~P. and {Cooray}, F.~R. and {Gupta}, N. and {Hayman}, D.~B. and {Jeganathan}, K. and {Koribalski}, B. and {Lee-Waddell}, K. and {Mahony}, Elizabeth K. and {Marvil}, J. and {McClure-Griffiths}, N.~M. and {Mirtschin}, P. and {Ng}, A. and {Pearce}, S. and {Phillips}, C. and {Voronkov}, M.~A.},
        title = "{The ASKAP Variables and Slow Transients (VAST) Pilot Survey}",
      journal = {\pasa},
         year = 2021,
        month = oct,
       volume = {38},
          eid = {e054},
        pages = {e054},
          doi = {10.1017/pasa.2021.44},
archivePrefix = {arXiv},
       eprint = {2108.06039},
 primaryClass = {astro-ph.HE},
       adsurl = {https://ui.adsabs.harvard.edu/abs/2021PASA...38...54M}
}

@ARTICLE{rose24,
       author = {{Rose}, Kovi and {Horesh}, Assaf and {Murphy}, Tara and {Kaplan}, David L. and {Sfaradi}, Itai and {Ryder}, Stuart D. and {Aloisi}, Robert J. and {Dobie}, Dougal and {Driessen}, Laura and {Fender}, Rob and {Green}, David A. and {Leung}, James K. and {Lenc}, Emil and {Qiu}, Hao and {Williams-Baldwin}, David},
        title = "{Late-time supernovae radio re-brightening in the VAST pilot survey}",
      journal = {\mnras},
         year = 2024,
        month = nov,
       volume = {534},
       number = {4},
        pages = {3853-3868},
          doi = {10.1093/mnras/stae2289},
archivePrefix = {arXiv},
       eprint = {2410.01375},
 primaryClass = {astro-ph.HE},
       adsurl = {https://ui.adsabs.harvard.edu/abs/2024MNRAS.534.3853R}
}

@ARTICLE{balasubramian21,
       author = {{Balasubramanian}, A. and {Corsi}, A. and {Polisensky}, E. and {Clarke}, T.~E. and {Kassim}, N.~E.},
        title = "{Radio Observations of SN2004dk with VLITE Confirm Late-time Rebrightening}",
      journal = {\apj},
         year = 2021,
        month = dec,
       volume = {923},
       number = {1},
          eid = {32},
        pages = {32},
          doi = {10.3847/1538-4357/ac2154},
archivePrefix = {arXiv},
       eprint = {2101.07348},
 primaryClass = {astro-ph.HE},
       adsurl = {https://ui.adsabs.harvard.edu/abs/2021ApJ...923...32B}
}

@ARTICLE{brethauer20,
       author = {{Brethauer}, Daniel and {Margutti}, Raffaella and {Milisavljevic}, Danny and {Bietenholz}, Michael},
        title = "{Six years of luminous X-ray emission from the strongly interacting type-Ib SN 2014C captured by Chandra and NuSTAR}",
      journal = {arXiv e-prints},
         year = 2020,
        month = dec,
          eid = {arXiv:2012.04081},
        pages = {arXiv:2012.04081},
          doi = {10.48550/arXiv.2012.04081},
archivePrefix = {arXiv},
       eprint = {2012.04081},
 primaryClass = {astro-ph.HE},
       adsurl = {https://ui.adsabs.harvard.edu/abs/2020arXiv201204081B}
}

@PHDTHESIS{briggs95,
       author = {{Briggs}, Daniel Shenon},
        title = "{High fidelity deconvolution of moderately resolved sources}",
       school = {New Mexico Institute of Mining and Technology},
         year = 1995,
        month = jan,
       adsurl = {https://ui.adsabs.harvard.edu/abs/1995PhDT.......238B}
}

@ARTICLE{casa22,
       author = {{CASA Team} and {Bean}, Ben and {Bhatnagar}, Sanjay and {Castro}, Sandra and {Donovan Meyer}, Jennifer and {Emonts}, Bjorn and {Garcia}, Enrique and {Garwood}, Robert and {Golap}, Kumar and {Gonzalez Villalba}, Justo and {Harris}, Pamela and {Hayashi}, Yohei and {Hoskins}, Josh and {Hsieh}, Mingyu and {Jagannathan}, Preshanth and {Kawasaki}, Wataru and {Keimpema}, Aard and {Kettenis}, Mark and {Lopez}, Jorge and {Marvil}, Joshua and {Masters}, Joseph and {McNichols}, Andrew and {Mehringer}, David and {Miel}, Renaud and {Moellenbrock}, George and {Montesino}, Federico and {Nakazato}, Takeshi and {Ott}, Juergen and {Petry}, Dirk and {Pokorny}, Martin and {Raba}, Ryan and {Rau}, Urvashi and {Schiebel}, Darrell and {Schweighart}, Neal and {Sekhar}, Srikrishna and {Shimada}, Kazuhiko and {Small}, Des and {Steeb}, Jan-Willem and {Sugimoto}, Kanako and {Suoranta}, Ville and {Tsutsumi}, Takahiro and {van Bemmel}, Ilse M. and {Verkouter}, Marjolein and {Wells}, Akeem and {Xiong}, Wei and {Szomoru}, Arpad and {Griffith}, Morgan and {Glendenning}, Brian and {Kern}, Jeff},
        title = "{CASA, the Common Astronomy Software Applications for Radio Astronomy}",
      journal = {\pasp},
         year = 2022,
        month = nov,
       volume = {134},
       number = {1041},
          eid = {114501},
        pages = {114501},
          doi = {10.1088/1538-3873/ac9642},
archivePrefix = {arXiv},
       eprint = {2210.02276},
 primaryClass = {astro-ph.IM},
       adsurl = {https://ui.adsabs.harvard.edu/abs/2022PASP..134k4501C}
}

@ARTICLE{castor75,
       author = {{Castor}, J. and {McCray}, R. and {Weaver}, R.},
        title = "{Interstellar bubbles.}",
      journal = {\apjl},
         year = 1975,
        month = sep,
       volume = {200},
        pages = {L107-L110},
          doi = {10.1086/181908},
       adsurl = {https://ui.adsabs.harvard.edu/abs/1975ApJ...200L.107C}
}

@ARTICLE{cendes18,
       author = {{Cendes}, Y. and {Gaensler}, B.~M. and {Ng}, C. -Y. and {Zanardo}, G. and {Staveley-Smith}, L. and {Tzioumis}, A.~K.},
        title = "{The Reacceleration of the Shock Wave in the Radio Remnant of SN 1987A}",
      journal = {\apj},
         year = 2018,
        month = nov,
       volume = {867},
       number = {1},
          eid = {65},
        pages = {65},
          doi = {10.3847/1538-4357/aae261},
archivePrefix = {arXiv},
       eprint = {1809.02364},
 primaryClass = {astro-ph.SR},
       adsurl = {https://ui.adsabs.harvard.edu/abs/2018ApJ...867...65C}
}

@ARTICLE{chevalier82a,
       author = {{Chevalier}, R.~A.},
        title = "{Self-similar solutions for the interaction of stellar ejecta with an external medium.}",
      journal = {\apj},
         year = 1982,
        month = jul,
       volume = {258},
        pages = {790-797},
          doi = {10.1086/160126},
       adsurl = {https://ui.adsabs.harvard.edu/abs/1982ApJ...258..790C}
}

@ARTICLE{chevalier82,
       author = {{Chevalier}, R.~A.},
        title = "{The radio and X-ray emission from type II supernovae.}",
      journal = {\apj},
         year = 1982,
        month = aug,
       volume = {259},
        pages = {302-310},
          doi = {10.1086/160167},
       adsurl = {https://ui.adsabs.harvard.edu/abs/1982ApJ...259..302C}
}

@ARTICLE{chevalier98,
       author = {{Chevalier}, R.~A.},
        title = "{Synchrotron Self-Absorption in Radio Supernovae}",
      journal = {\apj},
         year = 1998,
        month = may,
       volume = {499},
       number = {2},
        pages = {810-819},
          doi = {10.1086/305676},
       adsurl = {https://ui.adsabs.harvard.edu/abs/1998ApJ...499..810C}
}

@ARTICLE{chevalier06a,
       author = {{Chevalier}, Roger A. and {Fransson}, Claes and {Nymark}, Tanja K.},
        title = "{Radio and X-Ray Emission as Probes of Type IIP Supernovae and Red Supergiant Mass Loss}",
      journal = {\apj},
         year = 2006,
        month = apr,
       volume = {641},
       number = {2},
        pages = {1029-1038},
          doi = {10.1086/500528},
archivePrefix = {arXiv},
       eprint = {astro-ph/0509468},
 primaryClass = {astro-ph},
       adsurl = {https://ui.adsabs.harvard.edu/abs/2006ApJ...641.1029C}
}

@ARTICLE{chevalier06b,
       author = {{Chevalier}, Roger A. and {Fransson}, Claes},
        title = "{Circumstellar Emission from Type Ib and Ic Supernovae}",
      journal = {\apj},
         year = 2006,
        month = nov,
       volume = {651},
       number = {1},
        pages = {381-391},
          doi = {10.1086/507606},
archivePrefix = {arXiv},
       eprint = {astro-ph/0607196},
 primaryClass = {astro-ph},
       adsurl = {https://ui.adsabs.harvard.edu/abs/2006ApJ...651..381C}
}

@ARTICLE{chevalier10,
       author = {{Chevalier}, Roger A. and {Soderberg}, Alicia M.},
        title = "{Type IIb Supernovae with Compact and Extended Progenitors}",
      journal = {\apjl},
         year = 2010,
        month = mar,
       volume = {711},
       number = {1},
        pages = {L40-L43},
          doi = {10.1088/2041-8205/711/1/L40},
archivePrefix = {arXiv},
       eprint = {0911.3408},
 primaryClass = {astro-ph.HE},
       adsurl = {https://ui.adsabs.harvard.edu/abs/2010ApJ...711L..40C}
}

@INCOLLECTION{chevalier17,
       author = {{Chevalier}, Roger A. and {Fransson}, Claes},
        title = "{Thermal and Non-thermal Emission from Circumstellar Interaction}",
    booktitle = {Handbook of Supernovae},
         year = 2017,
       editor = {{Alsabti}, Athem W. and {Murdin}, Paul},
        pages = {875},
          doi = {10.1007/978-3-319-21846-5_34},
       adsurl = {https://ui.adsabs.harvard.edu/abs/2017hsn..book..875C}
}

@ARTICLE{claeys11,
       author = {{Claeys}, J.~S.~W. and {de Mink}, S.~E. and {Pols}, O.~R. and {Eldridge}, J.~J. and {Baes}, M.},
        title = "{Binary progenitor models of type IIb supernovae}",
      journal = {\aap},
         year = 2011,
        month = apr,
       volume = {528},
          eid = {A131},
        pages = {A131},
          doi = {10.1051/0004-6361/201015410},
archivePrefix = {arXiv},
       eprint = {1102.1732},
 primaryClass = {astro-ph.SR},
       adsurl = {https://ui.adsabs.harvard.edu/abs/2011A&A...528A.131C}
}

@ARTICLE{clocchiatti01,
       author = {{Clocchiatti}, A. and {Prieto}, J.~L.},
        title = "{Supernova 2001ig in NGC 7424.}",
      journal = {\iaucirc},
         year = 2001,
        month = dec,
       volume = {7781},
        pages = {2},
       adsurl = {https://ui.adsabs.harvard.edu/abs/2001IAUC.7781....2C}
}

@ARTICLE{clocchiatti02,
       author = {{Clocchiatti}, A.},
        title = "{Supernova 2001ig in NGC 7424}",
      journal = {\iaucirc},
         year = 2002,
        month = jan,
       volume = {7793},
        pages = {2},
       adsurl = {https://ui.adsabs.harvard.edu/abs/2002IAUC.7793....2C}
}

@ARTICLE{cox72,
       author = {{Cox}, Donald P.},
        title = "{Cooling and Evolution of a Supernova Remnant}",
      journal = {\apj},
         year = 1972,
        month = nov,
       volume = {178},
        pages = {159-168},
          doi = {10.1086/151775},
       adsurl = {https://ui.adsabs.harvard.edu/abs/1972ApJ...178..159C}
}

@ARTICLE{dessart11,
       author = {{Dessart}, Luc and {Hillier}, D. John and {Livne}, Eli and {Yoon}, Sung-Chul and {Woosley}, Stan and {Waldman}, Roni and {Langer}, Norbert},
        title = "{Core-collapse explosions of Wolf-Rayet stars and the connection to Type IIb/Ib/Ic supernovae}",
      journal = {\mnras},
         year = 2011,
        month = jul,
       volume = {414},
       number = {4},
        pages = {2985-3005},
          doi = {10.1111/j.1365-2966.2011.18598.x},
archivePrefix = {arXiv},
       eprint = {1102.5160},
 primaryClass = {astro-ph.SR},
       adsurl = {https://ui.adsabs.harvard.edu/abs/2011MNRAS.414.2985D}
}

@ARTICLE{evans01,
       author = {{Evans}, R.~O. and {White}, B. and {Bembrick}, C.},
        title = "{Supernova 2001ig in NGC 7424}",
      journal = {\iaucirc},
         year = 2001,
        month = dec,
       volume = {7772},
        pages = {1},
       adsurl = {https://ui.adsabs.harvard.edu/abs/2001IAUC.7772....1E}
}

@ARTICLE{filippenko88,
       author = {{Filippenko}, Alexei V.},
        title = "{Supernova 1987K: Type II in Youth, Type Ib in Old Age}",
      journal = {\aj},
         year = 1988,
        month = dec,
       volume = {96},
        pages = {1941},
          doi = {10.1086/114940},
       adsurl = {https://ui.adsabs.harvard.edu/abs/1988AJ.....96.1941F}
}

@ARTICLE{filippenko97,
       author = {{Filippenko}, Alexei V.},
        title = "{Optical Spectra of Supernovae}",
      journal = {\araa},
         year = 1997,
        month = jan,
       volume = {35},
        pages = {309-355},
          doi = {10.1146/annurev.astro.35.1.309},
       adsurl = {https://ui.adsabs.harvard.edu/abs/1997ARA&A..35..309F}
}

@ARTICLE{filippenko02,
       author = {{Filippenko}, A.~V. and {Chornock}, R.},
        title = "{Supernovae 2001ig, 2002eg, 2002ew, 2002gc}",
      journal = {\iaucirc},
         year = 2002,
        month = oct,
       volume = {7988},
        pages = {3},
       adsurl = {https://ui.adsabs.harvard.edu/abs/2002IAUC.7988....3F}
}

@ARTICLE{fox14,
       author = {{Fox}, Ori D. and {Azalee Bostroem}, K. and {Van Dyk}, Schuyler D. and {Filippenko}, Alexei V. and {Fransson}, Claes and {Matheson}, Thomas and {Cenko}, S. Bradley and {Chandra}, Poonam and {Dwarkadas}, Vikram and {Li}, Weidong and {Parker}, Alex H. and {Smith}, Nathan},
        title = "{Uncovering the Putative B-star Binary Companion of the SN 1993J Progenitor}",
      journal = {\apj},
         year = 2014,
        month = jul,
       volume = {790},
       number = {1},
          eid = {17},
        pages = {17},
          doi = {10.1088/0004-637X/790/1/17},
archivePrefix = {arXiv},
       eprint = {1405.4863},
 primaryClass = {astro-ph.HE},
       adsurl = {https://ui.adsabs.harvard.edu/abs/2014ApJ...790...17F}
}

@ARTICLE{gilkis22,
       author = {{Gilkis}, Avishai and {Arcavi}, Iair},
        title = "{How much hydrogen is in Type Ib and IIb supernova progenitors?}",
      journal = {\mnras},
         year = 2022,
        month = mar,
       volume = {511},
       number = {1},
        pages = {691-712},
          doi = {10.1093/mnras/stac088},
archivePrefix = {arXiv},
       eprint = {2111.04432},
 primaryClass = {astro-ph.SR},
       adsurl = {https://ui.adsabs.harvard.edu/abs/2022MNRAS.511..691G}
}

@ARTICLE{granot02,
       author = {{Granot}, Jonathan and {Panaitescu}, Alin and {Kumar}, Pawan and {Woosley}, Stan E.},
        title = "{Off-Axis Afterglow Emission from Jetted Gamma-Ray Bursts}",
      journal = {\apjl},
         year = 2002,
        month = may,
       volume = {570},
       number = {2},
        pages = {L61-L64},
          doi = {10.1086/340991},
archivePrefix = {arXiv},
       eprint = {astro-ph/0201322},
 primaryClass = {astro-ph},
       adsurl = {https://ui.adsabs.harvard.edu/abs/2002ApJ...570L..61G}
}

@ARTICLE{grupe16,
       author = {{Grupe}, Dirk and {Brown}, Peter and {Dong}, Subo and {Shappee}, B.~J. and {Holoien}, Tom and {Stanek}, Kris and {Prieto}, Jose L. and {Margutti}, Raffaella},
        title = "{X-ray and UV detections of the supernova candidate ASASSN 16fp (AT 2016coi) with Swift}",
      journal = {The Astronomer's Telegram},
         year = 2016,
        month = may,
       volume = {9088},
        pages = {1},
       adsurl = {https://ui.adsabs.harvard.edu/abs/2016ATel.9088....1G}
}

@ARTICLE{Gupta24,
       author = {{Gupta}, Siddhartha and {Caprioli}, Damiano and {Spitkovsky}, Anatoly},
        title = "{Electron Acceleration at Quasi-parallel Non-relativistic Shocks: A 1D Kinetic Survey}",
      journal = {arXiv e-prints},
         year = 2024,
        month = aug,
          eid = {arXiv:2408.16071},
        pages = {arXiv:2408.16071},
          doi = {10.48550/arXiv.2408.16071},
archivePrefix = {arXiv},
       eprint = {2408.16071},
 primaryClass = {astro-ph.HE},
       adsurl = {https://ui.adsabs.harvard.edu/abs/2024arXiv240816071G}
}

@ARTICLE{harris20,
       author = {{Harris}, C.~E. and {Nugent}, P.~E.},
        title = "{Outside the Wall: Hydrodynamics of Type I Supernovae Interacting with a Partially Swept-up Circumstellar Medium}",
      journal = {\apj},
         year = 2020,
        month = may,
       volume = {894},
       number = {2},
          eid = {122},
        pages = {122},
          doi = {10.3847/1538-4357/ab879e},
archivePrefix = {arXiv},
       eprint = {2004.03612},
 primaryClass = {astro-ph.HE},
       adsurl = {https://ui.adsabs.harvard.edu/abs/2020ApJ...894..122H}
}

@ARTICLE{horesh13,
       author = {{Horesh}, Assaf and {Stockdale}, Christopher and {Fox}, Derek B. and {Frail}, Dale A. and {Carpenter}, John and {Kulkarni}, S.~R. and {Ofek}, Eran O. and {Gal-Yam}, Avishay and {Kasliwal}, Mansi M. and {Arcavi}, Iair and {Quimby}, Robert and {Cenko}, S. Bradley and {Nugent}, Peter E. and {Bloom}, Joshua S. and {Law}, Nicholas M. and {Poznanski}, Dovi and {Gorbikov}, Evgeny and {Polishook}, David and {Yaron}, Ofer and {Ryder}, Stuart and {Weiler}, Kurt W. and {Bauer}, Franz and {Van Dyk}, Schuyler D. and {Immler}, Stefan and {Panagia}, Nino and {Pooley}, Dave and {Kassim}, Namir},
        title = "{An early and comprehensive millimetre and centimetre wave and X-ray study of SN 2011dh: a non-equipartition blast wave expanding into a massive stellar wind}",
      journal = {\mnras},
         year = 2013,
        month = dec,
       volume = {436},
       number = {2},
        pages = {1258-1267},
          doi = {10.1093/mnras/stt1645},
archivePrefix = {arXiv},
       eprint = {1209.1102},
 primaryClass = {astro-ph.CO},
       adsurl = {https://ui.adsabs.harvard.edu/abs/2013MNRAS.436.1258H}
}

@ARTICLE{hwang09,
       author = {{Hwang}, Una and {Laming}, J. Martin},
        title = "{The Circumstellar Medium of Cassiopeia a Inferred from the Outer Ejecta Knot Properties}",
      journal = {\apj},
         year = 2009,
        month = sep,
       volume = {703},
       number = {1},
        pages = {883-893},
          doi = {10.1088/0004-637X/703/1/883},
archivePrefix = {arXiv},
       eprint = {0907.5177},
 primaryClass = {astro-ph.HE},
       adsurl = {https://ui.adsabs.harvard.edu/abs/2009ApJ...703..883H}
}

@ARTICLE{idik24,
       author = {{Ibik}, Adaeze L. and {Drout}, Maria R. and {Margutti}, Raffaela and {Matthews}, David and {Villar}, V. Ashley and {Berger}, Edo and {Chornock}, Ryan and {Alexander}, Kate D. and {Eftekhari}, Tarraneh and {Laskar}, Tanmoy and {Lunnan}, Ragnhild and {Foley}, Ryan J. and {Jones}, David and {Milisavljevic}, Dan and {Rest}, Armin and {Scolnic}, Daniel and {Williams}, Peter K.~G.},
        title = "{PS1-11aop: Probing the Mass Loss History of a Luminous Interacting Supernova Prior to its Final Eruption with Multi-wavelength Observations}",
      journal = {arXiv e-prints},
         year = 2024,
        month = oct,
          eid = {arXiv:2410.15140},
        pages = {arXiv:2410.15140},
          doi = {10.48550/arXiv.2410.15140},
archivePrefix = {arXiv},
       eprint = {2410.15140},
 primaryClass = {astro-ph.HE},
       adsurl = {https://ui.adsabs.harvard.edu/abs/2024arXiv241015140I}
}

@ARTICLE{kamble16,
       author = {{Kamble}, Atish and {Margutti}, Raffaella and {Soderberg}, Alicia M. and {Chakraborti}, Sayan and {Fransson}, Claes and {Chevalier}, Roger and {Powell}, Diana and {Milisavljevic}, Dan and {Parrent}, Jerod and {Bietenholz}, Michael},
        title = "{Progenitors of Type IIB Supernovae in the Light of Radio and X-Rays from SN 2013DF}",
      journal = {\apj},
         year = 2016,
        month = feb,
       volume = {818},
       number = {2},
          eid = {111},
        pages = {111},
          doi = {10.3847/0004-637X/818/2/111},
archivePrefix = {arXiv},
       eprint = {1504.07988},
 primaryClass = {astro-ph.HE},
       adsurl = {https://ui.adsabs.harvard.edu/abs/2016ApJ...818..111K}
}

@BOOK{kippenhahn90,
       author = {{Kippenhahn}, Rudolf and {Weigert}, Alfred},
        title = "{Stellar Structure and Evolution}",
         year = 1990,
       adsurl = {https://ui.adsabs.harvard.edu/abs/1990sse..book.....K}
}

@ARTICLE{krause08,
       author = {{Krause}, Oliver and {Birkmann}, Stephan M. and {Usuda}, Tomonori and {Hattori}, Takashi and {Goto}, Miwa and {Rieke}, George H. and {Misselt}, Karl A.},
        title = "{The Cassiopeia A Supernova Was of Type IIb}",
      journal = {Science},
         year = 2008,
        month = may,
       volume = {320},
       number = {5880},
        pages = {1195},
          doi = {10.1126/science.1155788},
archivePrefix = {arXiv},
       eprint = {0805.4557},
 primaryClass = {astro-ph},
       adsurl = {https://ui.adsabs.harvard.edu/abs/2008Sci...320.1195K}
}

@ARTICLE{lau17,
       author = {{Lau}, R.~M. and {Hankins}, M.~J. and {Sch{\"o}del}, R. and {Sanchez-Bermudez}, J. and {Moffat}, A.~F.~J. and {Ressler}, M.~E.},
        title = "{Stagnant Shells in the Vicinity of the Dusty Wolf-Rayet-OB Binary WR 112}",
      journal = {\apjl},
         year = 2017,
        month = feb,
       volume = {835},
       number = {2},
          eid = {L31},
        pages = {L31},
          doi = {10.3847/2041-8213/835/2/L31},
archivePrefix = {arXiv},
       eprint = {1612.05650},
 primaryClass = {astro-ph.SR},
       adsurl = {https://ui.adsabs.harvard.edu/abs/2017ApJ...835L..31L}
}

@ARTICLE{mcconnell12,
       author = {{McConnell}, D. and {Sadler}, E.~M. and {Murphy}, T. and {Ekers}, R.~D.},
        title = "{ATPMN: accurate positions and flux densities at 5 and 8 GHz for 8385 sources from the PMN survey}",
      journal = {\mnras},
         year = 2012,
        month = may,
       volume = {422},
       number = {2},
        pages = {1527-1545},
          doi = {10.1111/j.1365-2966.2012.20726.x},
archivePrefix = {arXiv},
       eprint = {1202.2625},
 primaryClass = {astro-ph.CO},
       adsurl = {https://ui.adsabs.harvard.edu/abs/2012MNRAS.422.1527M}
}

@ARTICLE{maeda23,
       author = {{Maeda}, Keiichi and {Michiyama}, Tomonari and {Chandra}, Poonam and {Ryder}, Stuart and {Kuncarayakti}, Hanindyo and {Hiramatsu}, Daichi and {Imanishi}, Masatoshi},
        title = "{Resurrection of Type IIL Supernova 2018ivc: Implications for a Binary Evolution Sequence Connecting Hydrogen-rich and Hydrogen-poor Progenitors}",
      journal = {\apjl},
         year = 2023,
        month = mar,
       volume = {945},
       number = {1},
          eid = {L3},
        pages = {L3},
          doi = {10.3847/2041-8213/acb25e},
archivePrefix = {arXiv},
       eprint = {2301.07357},
 primaryClass = {astro-ph.HE},
       adsurl = {https://ui.adsabs.harvard.edu/abs/2023ApJ...945L...3M}
}

@ARTICLE{marchenko02,
       author = {{Marchenko}, S.~V. and {Moffat}, A.~F.~J. and {Vacca}, W.~D. and {C{\^o}t{\'e}}, S. and {Doyon}, R.},
        title = "{Massive Binary WR 112 and Properties of Wolf-Rayet Dust}",
      journal = {\apjl},
         year = 2002,
        month = jan,
       volume = {565},
       number = {1},
        pages = {L59-L62},
          doi = {10.1086/339138},
archivePrefix = {arXiv},
       eprint = {astro-ph/0112403},
 primaryClass = {astro-ph},
       adsurl = {https://ui.adsabs.harvard.edu/abs/2002ApJ...565L..59M}
}

@ARTICLE{margutti17,
       author = {{Margutti}, Raffaella and {Kamble}, A. and {Milisavljevic}, D. and {Zapartas}, E. and {de Mink}, S.~E. and {Drout}, M. and {Chornock}, R. and {Risaliti}, G. and {Zauderer}, B.~A. and {Bietenholz}, M. and {Cantiello}, M. and {Chakraborti}, S. and {Chomiuk}, L. and {Fong}, W. and {Grefenstette}, B. and {Guidorzi}, C. and {Kirshner}, R. and {Parrent}, J.~T. and {Patnaude}, D. and {Soderberg}, A.~M. and {Gehrels}, N.~C. and {Harrison}, F.},
        title = "{Ejection of the Massive Hydrogen-rich Envelope Timed with the Collapse of the Stripped SN 2014C}",
      journal = {\apj},
         year = 2017,
        month = feb,
       volume = {835},
       number = {2},
          eid = {140},
        pages = {140},
          doi = {10.3847/1538-4357/835/2/140},
archivePrefix = {arXiv},
       eprint = {1601.06806},
 primaryClass = {astro-ph.HE},
       adsurl = {https://ui.adsabs.harvard.edu/abs/2017ApJ...835..140M}
}

@ARTICLE{massardi11,
       author = {{Massardi}, Marcella and {Bonaldi}, Anna and {Bonavera}, Laura and {L{\'o}pez-Caniego}, Marcos and {de Zotti}, Gianfranco and {Ekers}, Ronald D.},
        title = "{The Planck-ATCA Co-eval Observations project: the bright sample}",
      journal = {\mnras},
         year = 2011,
        month = aug,
       volume = {415},
       number = {2},
        pages = {1597-1610},
          doi = {10.1111/j.1365-2966.2011.18802.x},
archivePrefix = {arXiv},
       eprint = {1101.0225},
 primaryClass = {astro-ph.CO},
       adsurl = {https://ui.adsabs.harvard.edu/abs/2011MNRAS.415.1597M}
}

@ARTICLE{matzner99,
       author = {{Matzner}, Christopher D. and {McKee}, Christopher F.},
        title = "{The Expulsion of Stellar Envelopes in Core-Collapse Supernovae}",
      journal = {\apj},
         year = 1999,
        month = jan,
       volume = {510},
       number = {1},
        pages = {379-403},
          doi = {10.1086/306571},
archivePrefix = {arXiv},
       eprint = {astro-ph/9807046},
 primaryClass = {astro-ph},
       adsurl = {https://ui.adsabs.harvard.edu/abs/1999ApJ...510..379M}
}

@ARTICLE{maund07,
       author = {{Maund}, Justyn R. and {Wheeler}, J. Craig and {Patat}, Ferdinando and {Wang}, Lifan and {Baade}, Dietrich and {H{\"o}flich}, Peter A.},
        title = "{Spectropolarimetry of the Type IIb Supernova 2001ig}",
      journal = {\apj},
         year = 2007,
        month = dec,
       volume = {671},
       number = {2},
        pages = {1944-1958},
          doi = {10.1086/523261},
archivePrefix = {arXiv},
       eprint = {0709.1487},
 primaryClass = {astro-ph},
       adsurl = {https://ui.adsabs.harvard.edu/abs/2007ApJ...671.1944M}
}

@ARTICLE{meunier13,
       author = {{Meunier}, C. and {Bauer}, F.~E. and {Dwarkadas}, V.~V. and {Koribalski}, B. and {Emonts}, B. and {Hunstead}, R.~W. and {Campbell-Wilson}, D. and {Stockdale}, C. and {Tingay}, S.~J.},
        title = "{Performing a stellar autopsy using the radio-bright remnant of SN 1996cr}",
      journal = {\mnras},
         year = 2013,
        month = may,
       volume = {431},
       number = {3},
        pages = {2453-2463},
          doi = {10.1093/mnras/stt340},
archivePrefix = {arXiv},
       eprint = {1302.5432},
 primaryClass = {astro-ph.HE},
       adsurl = {https://ui.adsabs.harvard.edu/abs/2013MNRAS.431.2453M}
}

@ARTICLE{milisavijevic15,
       author = {{Milisavljevic}, D. and {Margutti}, R. and {Kamble}, A. and {Patnaude}, D.~J. and {Raymond}, J.~C. and {Eldridge}, J.~J. and {Fong}, W. and {Bietenholz}, M. and {Challis}, P. and {Chornock}, R. and {Drout}, M.~R. and {Fransson}, C. and {Fesen}, R.~A. and {Grindlay}, J.~E. and {Kirshner}, R.~P. and {Lunnan}, R. and {Mackey}, J. and {Miller}, G.~F. and {Parrent}, J.~T. and {Sanders}, N.~E. and {Soderberg}, A.~M. and {Zauderer}, B.~A.},
        title = "{Metamorphosis of SN 2014C: Delayed Interaction between a Hydrogen Poor Core-collapse Supernova and a Nearby Circumstellar Shell}",
      journal = {\apj},
         year = 2015,
        month = dec,
       volume = {815},
       number = {2},
          eid = {120},
        pages = {120},
          doi = {10.1088/0004-637X/815/2/120},
archivePrefix = {arXiv},
       eprint = {1511.01907},
 primaryClass = {astro-ph.HE},
       adsurl = {https://ui.adsabs.harvard.edu/abs/2015ApJ...815..120M}
}

@ARTICLE{monnier99,
       author = {{Monnier}, J.~D. and {Tuthill}, P.~G. and {Danchi}, W.~C.},
        title = "{Pinwheel Nebula around WR 98A}",
      journal = {\apjl},
         year = 1999,
        month = nov,
       volume = {525},
       number = {2},
        pages = {L97-L100},
          doi = {10.1086/312352},
archivePrefix = {arXiv},
       eprint = {astro-ph/9909282},
 primaryClass = {astro-ph},
       adsurl = {https://ui.adsabs.harvard.edu/abs/1999ApJ...525L..97M}
}

@ARTICLE{nomoto93,
       author = {{Nomoto}, K. and {Suzuki}, T. and {Shigeyama}, T. and {Kumagai}, S. and {Yamaoka}, H. and {Saio}, H.},
        title = "{A type IIb model for supernova 1993J}",
      journal = {\nat},
         year = 1993,
        month = aug,
       volume = {364},
       number = {6437},
        pages = {507-509},
          doi = {10.1038/364507a0},
       adsurl = {https://ui.adsabs.harvard.edu/abs/1993Natur.364..507N}
}

@ARTICLE{nyholm17,
       author = {{Nyholm}, A. and {Sollerman}, J. and {Taddia}, F. and {Fremling}, C. and {Moriya}, T.~J. and {Ofek}, E.~O. and {Gal-Yam}, A. and {De Cia}, A. and {Roy}, R. and {Kasliwal}, M.~M. and {Cao}, Y. and {Nugent}, P.~E. and {Masci}, F.~J.},
        title = "{The bumpy light curve of Type IIn supernova iPTF13z over 3 years}",
      journal = {\aap},
         year = 2017,
        month = aug,
       volume = {605},
          eid = {A6},
        pages = {A6},
          doi = {10.1051/0004-6361/201629906},
archivePrefix = {arXiv},
       eprint = {1703.09679},
 primaryClass = {astro-ph.SR},
       adsurl = {https://ui.adsabs.harvard.edu/abs/2017A&A...605A...6N}
}

@ARTICLE{paczynski01,
       author = {{Paczynski}, Bohdan},
        title = "{Gamma-Ray Bursts at Low Redshift}",
      journal = {\actaa},
         year = 2001,
        month = mar,
       volume = {51},
        pages = {1-4},
          doi = {10.48550/arXiv.astro-ph/0103384},
archivePrefix = {arXiv},
       eprint = {astro-ph/0103384},
 primaryClass = {astro-ph},
       adsurl = {https://ui.adsabs.harvard.edu/abs/2001AcA....51....1P}
}

@ARTICLE{phillips01,
       author = {{Phillips}, M.~M. and {Suntzeff}, N.~B. and {Krisciunas}, K. and {Carlberg}, R. and {Gladders}, M. and {Barrientos}, F. and {Matheson}, T. and {Jha}, S.},
        title = "{Supernova 2001ig in NGC 7424}",
      journal = {\iaucirc},
         year = 2001,
        month = dec,
       volume = {7772},
        pages = {2},
       adsurl = {https://ui.adsabs.harvard.edu/abs/2001IAUC.7772....2P}
}

@ARTICLE{podsiadlowski93,
       author = {{Podsiadlowski}, Ph. and {Hsu}, J.~J.~L. and {Joss}, P.~C. and {Ross}, R.~R.},
        title = "{The progenitor of supernova 1993J: a stripped supergiant in a binary system?}",
      journal = {\nat},
         year = 1993,
        month = aug,
       volume = {364},
       number = {6437},
        pages = {509-511},
          doi = {10.1038/364509a0},
       adsurl = {https://ui.adsabs.harvard.edu/abs/1993Natur.364..509P}
}

@ARTICLE{pooley19,
       author = {{Pooley}, David and {Wheeler}, J. Craig and {Vink{\'o}}, Jozsef and {Dwarkadas}, Vikram V. and {Szalai}, Tamas and {Silverman}, Jeffrey M. and {Griesel}, Madelaine and {McCullough}, Molly and {Marion}, G.~H. and {MacQueen}, Phillip},
        title = "{Interaction of SN Ib 2004dk with a Previously Expelled Envelope}",
      journal = {\apj},
         year = 2019,
        month = oct,
       volume = {883},
       number = {2},
          eid = {120},
        pages = {120},
          doi = {10.3847/1538-4357/ab3e36},
archivePrefix = {arXiv},
       eprint = {1910.06395},
 primaryClass = {astro-ph.HE},
       adsurl = {https://ui.adsabs.harvard.edu/abs/2019ApJ...883..120P}
}

@ARTICLE{quataert12,
       author = {{Quataert}, E. and {Shiode}, J.},
        title = "{Wave-driven mass loss in the last year of stellar evolution: setting the stage for the most luminous core-collapse supernovae}",
      journal = {\mnras},
         year = 2012,
        month = jun,
       volume = {423},
       number = {1},
        pages = {L92-L96},
          doi = {10.1111/j.1745-3933.2012.01264.x},
archivePrefix = {arXiv},
       eprint = {1202.5036},
 primaryClass = {astro-ph.SR},
       adsurl = {https://ui.adsabs.harvard.edu/abs/2012MNRAS.423L..92Q}
}

@ARTICLE{ryder04,
       author = {{Ryder}, Stuart D. and {Sadler}, Elaine M. and {Subrahmanyan}, Ravi and {Weiler}, Kurt W. and {Panagia}, Nino and {Stockdale}, Christopher},
        title = "{Modulations in the radio light curve of the Type IIb supernova 2001ig: evidence for a Wolf-Rayet binary progenitor?}",
      journal = {\mnras},
         year = 2004,
        month = apr,
       volume = {349},
       number = {3},
        pages = {1093-1100},
          doi = {10.1111/j.1365-2966.2004.07589.x},
archivePrefix = {arXiv},
       eprint = {astro-ph/0401135},
 primaryClass = {astro-ph},
       adsurl = {https://ui.adsabs.harvard.edu/abs/2004MNRAS.349.1093R}
}

@ARTICLE{ryder06,
       author = {{Ryder}, Stuart D. and {Murrowood}, Clair E. and {Stathakis}, Raylee A.},
        title = "{A post-mortem investigation of the Type IIb supernova 2001ig}",
      journal = {\mnras},
         year = 2006,
        month = jun,
       volume = {369},
       number = {1},
        pages = {L32-L36},
          doi = {10.1111/j.1745-3933.2006.00168.x},
archivePrefix = {arXiv},
       eprint = {astro-ph/0603336},
 primaryClass = {astro-ph},
       adsurl = {https://ui.adsabs.harvard.edu/abs/2006MNRAS.369L..32R}
}

@ARTICLE{ryder18,
       author = {{Ryder}, Stuart D. and {Van Dyk}, Schuyler D. and {Fox}, Ori D. and {Zapartas}, Emmanouil and {de Mink}, Selma E. and {Smith}, Nathan and {Brunsden}, Emily and {Azalee Bostroem}, K. and {Filippenko}, Alexei V. and {Shivvers}, Isaac and {Zheng}, WeiKang},
        title = "{Ultraviolet Detection of the Binary Companion to the Type IIb SN 2001ig}",
      journal = {\apj},
         year = 2018,
        month = mar,
       volume = {856},
       number = {1},
          eid = {83},
        pages = {83},
          doi = {10.3847/1538-4357/aaaf1e},
archivePrefix = {arXiv},
       eprint = {1801.05125},
 primaryClass = {astro-ph.SR},
       adsurl = {https://ui.adsabs.harvard.edu/abs/2018ApJ...856...83R}
}

@ARTICLE{schwarz96,
       author = {{Schwarz}, D.~H. and {Pringle}, J.~E.},
        title = "{A self-colliding stellar wind model for SN 1979C}",
      journal = {\mnras},
         year = 1996,
        month = oct,
       volume = {282},
       number = {3},
        pages = {1018-1026},
          doi = {10.1093/mnras/282.3.1018},
       adsurl = {https://ui.adsabs.harvard.edu/abs/1996MNRAS.282.1018S}
}

@ARTICLE{silverman09,
       author = {{Silverman}, Jeffrey M. and {Mazzali}, Paolo and {Chornock}, Ryan and {Filippenko}, Alexei V. and {Clocchiatti}, Alejandro and {Phillips}, Mark M. and {Ganeshalingam}, Mohan and {Foley}, Ryan J.},
        title = "{Optical Spectroscopy of the Somewhat Peculiar Type IIb Supernova 2001ig}",
      journal = {\pasp},
         year = 2009,
        month = jul,
       volume = {121},
       number = {881},
        pages = {689},
          doi = {10.1086/603653},
archivePrefix = {arXiv},
       eprint = {0903.4179},
 primaryClass = {astro-ph.CO},
       adsurl = {https://ui.adsabs.harvard.edu/abs/2009PASP..121..689S}
}

@INCOLLECTION{slane17,
       author = {{Slane}, Patrick},
        title = "{Pulsar Wind Nebulae}",
    booktitle = {Handbook of Supernovae},
         year = 2017,
       editor = {{Alsabti}, Athem W. and {Murdin}, Paul},
        pages = {2159},
          doi = {10.1007/978-3-319-21846-5_95},
       adsurl = {https://ui.adsabs.harvard.edu/abs/2017hsn..book.2159S}
}

@ARTICLE{soderberg06,
       author = {{Soderberg}, A.~M. and {Chevalier}, R.~A. and {Kulkarni}, S.~R. and {Frail}, D.~A.},
        title = "{The Radio and X-Ray Luminous SN 2003bg and the Circumstellar Density Variations around Radio Supernovae}",
      journal = {\apj},
         year = 2006,
        month = nov,
       volume = {651},
       number = {2},
        pages = {1005-1018},
          doi = {10.1086/507571},
archivePrefix = {arXiv},
       eprint = {astro-ph/0512413},
 primaryClass = {astro-ph},
       adsurl = {https://ui.adsabs.harvard.edu/abs/2006ApJ...651.1005S}
}

@ARTICLE{soria24,
       author = {{Soria}, Roberto and {Cheng}, Siying and {Pakull}, Manfred W. and {Motch}, Christian and {Russell}, Thomas D.},
        title = "{A multiband look at ultraluminous X-ray sources in NGC 7424}",
      journal = {\mnras},
         year = 2024,
        month = apr,
       volume = {529},
       number = {2},
        pages = {1169-1186},
          doi = {10.1093/mnras/stae551},
archivePrefix = {arXiv},
       eprint = {2402.09512},
 primaryClass = {astro-ph.HE},
       adsurl = {https://ui.adsabs.harvard.edu/abs/2024MNRAS.529.1169S}
}

@ARTICLE{soulain18,
       author = {{Soulain}, A. and {Millour}, F. and {Lopez}, B. and {Matter}, A. and {Lagadec}, E. and {Carbillet}, M. and {La Camera}, A. and {Lamberts}, A. and {Langlois}, M. and {Milli}, J. and {Avenhaus}, H. and {Magnard}, Y. and {Roux}, A. and {Moulin}, T. and {Carle}, M. and {Sevin}, A. and {Martinez}, P. and {Abe}, L. and {Ramos}, J.},
        title = "{SPHERE view of Wolf-Rayet 104. Direct detection of the Pinwheel and the link with the nearby star}",
      journal = {\aap},
         year = 2018,
        month = oct,
       volume = {618},
          eid = {A108},
        pages = {A108},
          doi = {10.1051/0004-6361/201832817},
archivePrefix = {arXiv},
       eprint = {1806.08525},
 primaryClass = {astro-ph.SR},
       adsurl = {https://ui.adsabs.harvard.edu/abs/2018A&A...618A.108S}
}

@INCOLLECTION{sramek03,
       author = {{Sramek}, R.~A. and {Weiler}, K.~W.},
        title = "{Radio Supernovae}",
    booktitle = {Supernovae and Gamma-Ray Bursters},
         year = 2003,
       editor = {{Weiler}, K.},
       volume = {598},
        pages = {145-169},
          doi = {10.1007/3-540-45863-8_9},
       adsurl = {https://ui.adsabs.harvard.edu/abs/2003LNP...598..145S}
}

@ARTICLE{sravan19,
       author = {{Sravan}, Niharika and {Marchant}, Pablo and {Kalogera}, Vassiliki},
        title = "{Progenitors of Type IIb Supernovae. I. Evolutionary Pathways and Rates}",
      journal = {\apj},
         year = 2019,
        month = nov,
       volume = {885},
       number = {2},
          eid = {130},
        pages = {130},
          doi = {10.3847/1538-4357/ab4ad7},
archivePrefix = {arXiv},
       eprint = {1808.07580},
 primaryClass = {astro-ph.SR},
       adsurl = {https://ui.adsabs.harvard.edu/abs/2019ApJ...885..130S}
}

@ARTICLE{sravan20,
       author = {{Sravan}, Niharika and {Marchant}, Pablo and {Kalogera}, Vassiliki and {Milisavljevic}, Dan and {Margutti}, Raffaella},
        title = "{Progenitors of Type IIb Supernovae. II. Observable Properties}",
      journal = {\apj},
         year = 2020,
        month = nov,
       volume = {903},
       number = {1},
          eid = {70},
        pages = {70},
          doi = {10.3847/1538-4357/abb8d5},
archivePrefix = {arXiv},
       eprint = {2009.06405},
 primaryClass = {astro-ph.HE},
       adsurl = {https://ui.adsabs.harvard.edu/abs/2020ApJ...903...70S}
}

@ARTICLE{stockdale07,
       author = {{Stockdale}, Christopher J. and {Williams}, Christopher L. and {Weiler}, Kurt W. and {Panagia}, Nino and {Sramek}, Richard A. and {Van Dyk}, Schuyler D. and {Kelley}, Matthew T.},
        title = "{The Radio Evolution of SN 2001gd}",
      journal = {\apj},
         year = 2007,
        month = dec,
       volume = {671},
       number = {1},
        pages = {689-694},
          doi = {10.1086/522584},
archivePrefix = {arXiv},
       eprint = {0708.1026},
 primaryClass = {astro-ph},
       adsurl = {https://ui.adsabs.harvard.edu/abs/2007ApJ...671..689S}
}

@ARTICLE{stroh21,
       author = {{Stroh}, Michael C. and {Terreran}, Giacomo and {Coppejans}, Deanne L. and {Bright}, Joe S. and {Margutti}, Raffaella and {Bietenholz}, Michael F. and {De Colle}, Fabio and {DeMarchi}, Lindsay and {Duran}, Rodolfo Barniol and {Milisavljevic}, Danny and {Murase}, Kohta and {Paterson}, Kerry and {Williams}, Wendy L.},
        title = "{Luminous Late-time Radio Emission from Supernovae Detected by the Karl G. Jansky Very Large Array Sky Survey (VLASS)}",
      journal = {\apjl},
         year = 2021,
        month = dec,
       volume = {923},
       number = {2},
          eid = {L24},
        pages = {L24},
          doi = {10.3847/2041-8213/ac375e},
archivePrefix = {arXiv},
       eprint = {2106.09737},
 primaryClass = {astro-ph.HE},
       adsurl = {https://ui.adsabs.harvard.edu/abs/2021ApJ...923L..24S}
}

@ARTICLE{terreran19,
       author = {{Terreran}, G. and {Margutti}, R. and {Bersier}, D. and {Brimacombe}, J. and {Caprioli}, D. and {Challis}, P. and {Chornock}, R. and {Coppejans}, D.~L. and {Dong}, Subo and {Guidorzi}, C. and {Hurley}, K. and {Kirshner}, R. and {Migliori}, G. and {Milisavljevic}, D. and {Palmer}, D.~M. and {Prieto}, J.~L. and {Tomasella}, L. and {Marchant}, P. and {Pastorello}, A. and {Shappee}, B.~J. and {Stanek}, K.~Z. and {Stritzinger}, M.~D. and {Benetti}, S. and {Chen}, Ping and {DeMarchi}, L. and {Elias-Rosa}, N. and {Gall}, C. and {Harmanen}, J. and {Mattila}, S.},
        title = "{SN 2016coi (ASASSN-16fp): An Energetic H-stripped Core-collapse Supernova from a Massive Stellar Progenitor with Large Mass Loss}",
      journal = {\apj},
         year = 2019,
        month = oct,
       volume = {883},
       number = {2},
          eid = {147},
        pages = {147},
          doi = {10.3847/1538-4357/ab3e37},
archivePrefix = {arXiv},
       eprint = {1905.02226},
 primaryClass = {astro-ph.HE},
       adsurl = {https://ui.adsabs.harvard.edu/abs/2019ApJ...883..147T}
}

@ARTICLE{tuthill99,
       author = {{Tuthill}, Peter G. and {Monnier}, John D. and {Danchi}, William C.},
        title = "{A dusty pinwheel nebula around the massive star WR104}",
      journal = {\nat},
         year = 1999,
        month = apr,
       volume = {398},
       number = {6727},
        pages = {487-489},
          doi = {10.1038/19033},
archivePrefix = {arXiv},
       eprint = {astro-ph/9904092},
 primaryClass = {astro-ph},
       adsurl = {https://ui.adsabs.harvard.edu/abs/1999Natur.398..487T}
}

@ARTICLE{vinko17,
       author = {{Vinko}, J. and {Pooley}, D. and {Silverman}, J.~M. and {Wheeler}, J.~C. and {Szalai}, T. and {Kelly}, P. and {MacQueen}, P. and {Marion}, G.~H. and {S{\'a}rneczky}, K.},
        title = "{Searching for the Expelled Hydrogen Envelope in Type I Supernovae via Late-Time H{\ensuremath{\alpha}} Emission}",
      journal = {\apj},
         year = 2017,
        month = mar,
       volume = {837},
       number = {1},
          eid = {62},
        pages = {62},
          doi = {10.3847/1538-4357/aa607e},
archivePrefix = {arXiv},
       eprint = {1702.05143},
 primaryClass = {astro-ph.HE},
       adsurl = {https://ui.adsabs.harvard.edu/abs/2017ApJ...837...62V}
}

@INPROCEEDINGS{walder03,
       author = {{Walder}, Rolf and {Folini}, Doris},
        title = "{3D-hydrodynamics of colliding winds in massive binaries}",
    booktitle = {A Massive Star Odyssey: From Main Sequence to Supernova},
         year = 2003,
       editor = {{van der Hucht}, Karel and {Herrero}, Artemio and {Esteban}, C{\'e}sar},
       volume = {212},
        month = jan,
        pages = {139},
       adsurl = {https://ui.adsabs.harvard.edu/abs/2003IAUS..212..139W}
}

@ARTICLE{weaver77,
       author = {{Weaver}, R. and {McCray}, R. and {Castor}, J. and {Shapiro}, P. and {Moore}, R.},
        title = "{Interstellar bubbles. II. Structure and evolution.}",
      journal = {\apj},
         year = 1977,
        month = dec,
       volume = {218},
        pages = {377-395},
          doi = {10.1086/155692},
       adsurl = {https://ui.adsabs.harvard.edu/abs/1977ApJ...218..377W}
}

@ARTICLE{weiler86,
       author = {{Weiler}, K.~W. and {Sramek}, R.~A. and {Panagia}, N. and {van der Hulst}, J.~M. and {Salvati}, M.},
        title = "{Radio Supernovae}",
      journal = {\apj},
         year = 1986,
        month = feb,
       volume = {301},
        pages = {790},
          doi = {10.1086/163944},
       adsurl = {https://ui.adsabs.harvard.edu/abs/1986ApJ...301..790W}
}

@ARTICLE{weiler90,
       author = {{Weiler}, Kurt W. and {Panagia}, Nino and {Sramek}, Richard A.},
        title = "{Radio Emission from Supernovae. II. SN 1986J: A Different Kind of Type II}",
      journal = {\apj},
         year = 1990,
        month = dec,
       volume = {364},
        pages = {611},
          doi = {10.1086/169444},
       adsurl = {https://ui.adsabs.harvard.edu/abs/1990ApJ...364..611W}
}

@ARTICLE{weiler02,
       author = {{Weiler}, Kurt W. and {Panagia}, Nino and {Montes}, Marcos J. and {Sramek}, Richard A.},
        title = "{Radio Emission from Supernovae and Gamma-Ray Bursters}",
      journal = {\araa},
         year = 2002,
        month = jan,
       volume = {40},
        pages = {387-438},
          doi = {10.1146/annurev.astro.40.060401.093744},
       adsurl = {https://ui.adsabs.harvard.edu/abs/2002ARA&A..40..387W}
}

@ARTICLE{weiler07,
       author = {{Weiler}, Kurt W. and {Williams}, Christopher L. and {Panagia}, Nino and {Stockdale}, Christopher J. and {Kelley}, Matthew T. and {Sramek}, Richard A. and {Van Dyk}, Schuyler D. and {Marcaide}, J.~M.},
        title = "{Long-Term Radio Monitoring of SN 1993J}",
      journal = {\apj},
         year = 2007,
        month = dec,
       volume = {671},
       number = {2},
        pages = {1959-1980},
          doi = {10.1086/523258},
archivePrefix = {arXiv},
       eprint = {0709.1136},
 primaryClass = {astro-ph},
       adsurl = {https://ui.adsabs.harvard.edu/abs/2007ApJ...671.1959W}
}

@ARTICLE{wellons12,
       author = {{Wellons}, Sarah and {Soderberg}, Alicia M. and {Chevalier}, Roger A.},
        title = "{Radio Observations Reveal Unusual Circumstellar Environments for Some Type Ibc Supernova Progenitors}",
      journal = {\apj},
         year = 2012,
        month = jun,
       volume = {752},
       number = {1},
          eid = {17},
        pages = {17},
          doi = {10.1088/0004-637X/752/1/17},
archivePrefix = {arXiv},
       eprint = {1201.5120},
 primaryClass = {astro-ph.HE},
       adsurl = {https://ui.adsabs.harvard.edu/abs/2012ApJ...752...17W}
}

@ARTICLE{yoon17,
       author = {{Yoon}, Sung-Chul and {Dessart}, Luc and {Clocchiatti}, Alejandro},
        title = "{Type Ib and IIb Supernova Progenitors in Interacting Binary Systems}",
      journal = {\apj},
         year = 2017,
        month = may,
       volume = {840},
       number = {1},
          eid = {10},
        pages = {10},
          doi = {10.3847/1538-4357/aa6afe},
archivePrefix = {arXiv},
       eprint = {1701.02089},
 primaryClass = {astro-ph.SR},
       adsurl = {https://ui.adsabs.harvard.edu/abs/2017ApJ...840...10Y}
}

\appendix

\section{Corner plots of the fit parameters}

\begin{figure*}
  \begin{center}
    \includegraphics[width=\textwidth]{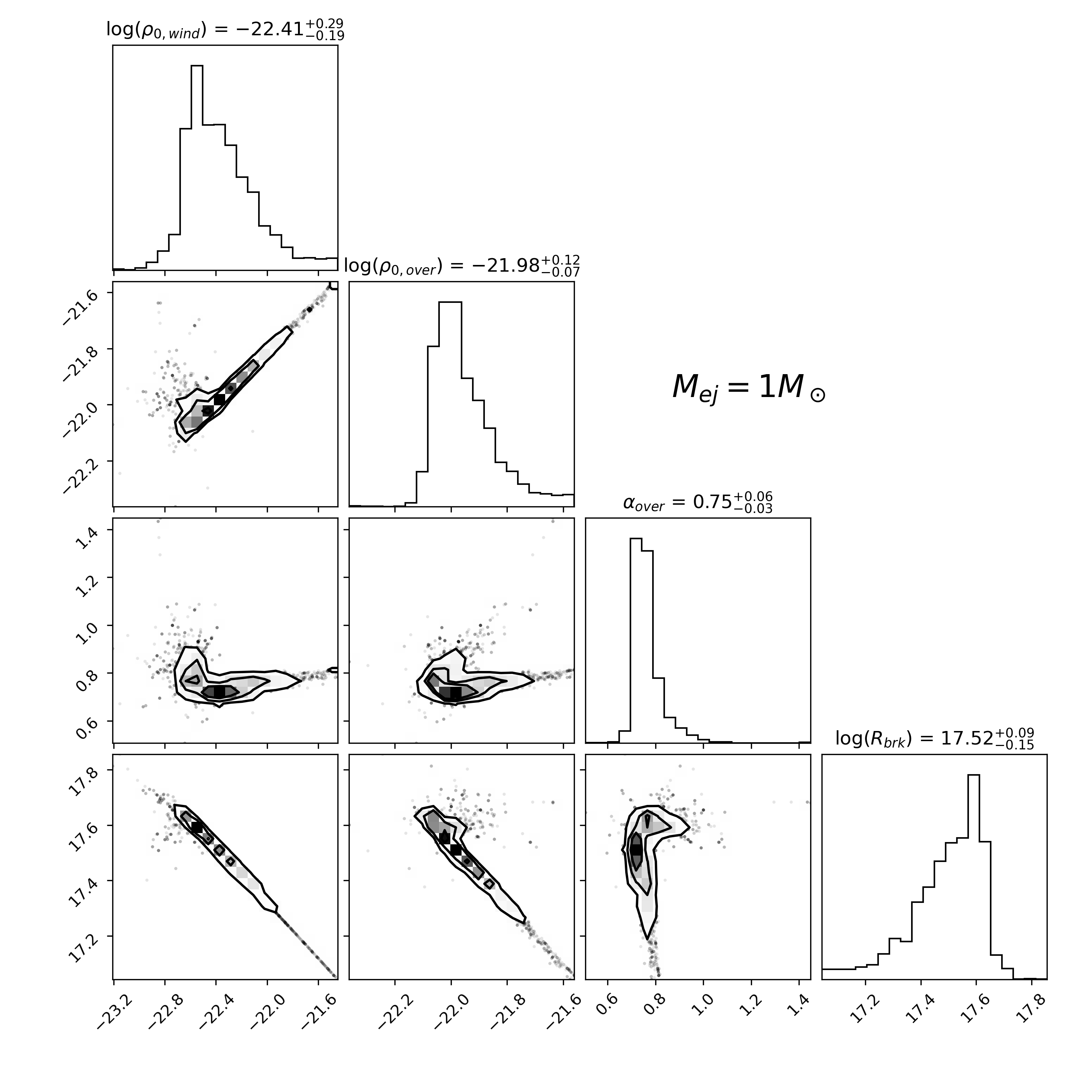}
  \end{center}
  \vspace{-2mm}
  \caption{
  Corner plot of the best-fitting model parameters and their uncertainties for the 1$M_\odot$ ejecta mass model, for the CSM the SN shock is propagating through (cgs units). $R_{\textrm{brk}}$ is the radius at which the shock encounters an overdensity. $\rho_{\textrm{0,wind}}$ and $\rho_{\textrm{0,over}}$ represent the CSM density immediately before and after $R_{\textrm{brk}}$. In our model, we have assumed that the CSM follows a wind density profile at $R < R_{\textrm{brk}}$ and a profile $\rho \propto R^{-\alpha_{\textrm{late}}}$ at $R > R_{\textrm{brk}}$. 
  }
  \vspace{-0.6cm}
  \label{fig:corner_plot1}
\end{figure*}

\begin{figure*}
  \begin{center}
    \includegraphics[width=\textwidth]{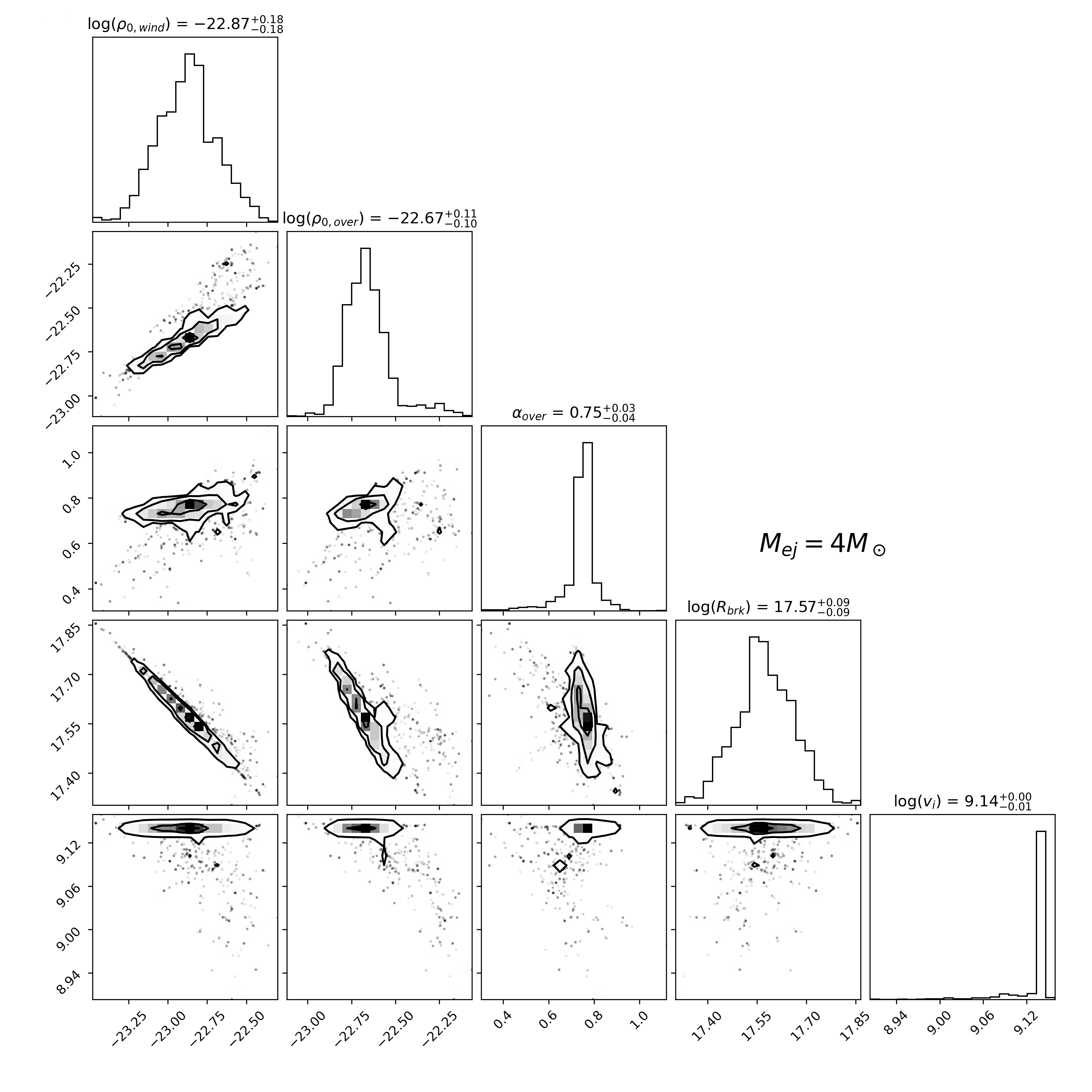}
  \end{center}
  \vspace{-2mm}
  \caption{
  As in Figure \ref{fig:corner_plot1}, for the 4$M_\odot$ ejecta mass model. 
  In this case, the initial shock velocity was also a fit parameter.
  }
  \vspace{-0.6cm}
  \label{fig:corner_plot2}
\end{figure*}

\end{document}